\documentclass{aastex61}

\newcommand\aastex{AAS\TeX}

\received{July 1, 2016}
\revised{September 27, 2016}
\accepted{\today}
\submitjournal{ApJ}

\shorttitle{\aastex\ sample article}
\shortauthors{Centeno}

\begin{document}

\title{On the weak field approximation for Ca {\sc II} 8542 \AA}

\correspondingauthor{Rebecca Centeno}
\email{rce@ucar.edu}

\author[0000-0002-1327-1278]{Rebecca Centeno}
\affil{High Altitude Observatory \\
3080 Center Green Dr.
Boulder, CO, USA}

\begin{abstract}
The weak field approximation (WFA) is a conceptually simple and computationally light method for inferring the magnetic field strength and its orientation in the Sun's atmosphere. In this work we study the validity and limitations of this tool when applied to full Stokes Ca {\sc ii} 8542 \AA\ profiles to extract information about the chromospheric magnetic field. We find that the range of validity of the WFA depends, amongst other things, on the component of the magnetic field that one is trying to infer. 

\noindent The retrieval of line-of-sight component of the chromospheric magnetic field from the core of the spectral line is reliable for field strengths up to $\sim 1200$~G, even when moderate velocity gradients are present.
The horizontal component, on the other hand, is suitably derived using the wing-core boundary of the spectral line, but typically yields systematic errors of $\geqslant 10\%
$. The effects of scattering polarization further compound the problem by rendering the transverse field inference problematic in quiet Sun areas, and for observing geometries within $30^{\circ}$ of the limb. Magneto-optical effects disproportionately challenge the determination of the magnetic field azimuth in the transverse plane, leading to errors of $\sim 10^{\circ}$.

\noindent Typical noise levels of $\sigma_{\rm n}=10^{-3}$ relative to the continuum intensity preclude the accurate retrieval of the transverse field strength and its azimuth below a threshold of  a few hundred gauss. Striving for a noise level of $\sigma_{\rm n}=10^{-4}$ significantly improves the diagnostic capability of the WFA with this spectral line, at which point the magnetic field inference becomes limited by systematic errors.

\end{abstract}

\section{Introduction} \label{sec:intro}

Our knowledge of the Sun relies, almost entirely, upon correctly interpreting the intensity and polarization of the light that leaves its atmosphere and hits our telescopes and detectors.

The solar chromosphere is a highly structured and very dynamic region of the Sun's atmosphere that witnesses the transition from a high-$\beta$ to a low-$\beta$ plasma regime, this is, it sits in between the dynamically-driven photosphere and the magnetically-dominated corona.
The low density environment of the chromosphere results in much less frequent particle collisions than in the photosphere, so the radiation is only weakly coupled to the local plasma conditions. Unlike in the Sun's photosphere, the simplifying assumption of local thermodynamic equilibrium (LTE) cannot generally be used to solve the radiative transfer problem. Thus, properly synthesizing a chromospheric spectral line requires solving, simultaneously and self-consistently, the radiative transfer (RT) and the statistical equilibrium (SE) equations (the latter govern the atomic state and the populations of the different atomic energy levels). This is a non-local, non-linear problem that involves an iterative scheme for the numerical integration of the RT equation in order to carry out the forward calculation alone. Interpreting observed spectra entails yet another layer of iterative cycles to solve the inverse problem, i.e., mapping the observations to the most likely physical properties of the atmosphere that generated them.

\noindent Geometric expansion effects make the chromospheric magnetic fields much weaker than their photospheric counterparts, so the polarization signatures that these fields imprint on spectral lines are also smaller. Measuring them with an adequate signal-to-noise ratio requires very sensitive instruments and extremely accurate calibration methods.

\noindent Thus, we confront two distinct challenges when interpreting chromospheric radiation and polarization. On the one hand, the measurement of the intrinsically weak polarization signals in the lines that are accessible to ground- and space-based observatories is a challenge to telescopes, instruments and detectors. On the other hand, the interpretation of the observed radiation relies on our knowledge of the physics that goes into the polarized radiative transfer and our capability to take on the increasingly large computational cost of its modeling. All of these issues render the interpretation of chromospheric radiation quite an arduous task.

There are three approaches that are typically used to interpret chromospheric spectra, ranging from fully solving the non-LTE radiative transfer problem to using the weak field approximation \citep[][]{jaime_review}. Listed in order of decreasing complexity and realism:

\begin{itemize}
\item Non-LTE spectral line inversions, which calculate the inverse mapping of the observed spectra to the physical properties of the solar atmosphere by solving the coupled system of the RT and the SE equations under certain simplifying assumptions (complete redistribution, 1D radiative transfer, no scattering polarization). To date, these methods have been almost exclusively applied to the Ca {\sc ii} IR triplet.
\item Milne-Eddington and slab inversions, which have mainly been used to interpret the spectral line intensity and polarization of the He {\sc i} 10830 \AA\ and D3 multiplets. These approximations can be used because of the peculiar formation mechanism of these particular lines, whose height of formation spans only a thin layer at the top of the chromosphere.
\item The weak field approximation (WFA), which relies on the Zeeman splitting to be smaller than the Doppler width of the spectral line, is a fast method to extract the magnetic field vector from spectropolarimetric observations. Chromospheric spectral lines tend to be rather broad, which expands the use of the WFA to larger field strengths than in the photosphere. However, the WFA works under some additional, strongly simplifying assumptions (constant velocity and magnetic field along the line-of-sight) that limit the breadth of its use.
\end{itemize}

Despite the difficulties and/or caveats of the afore-mentioned approaches, determining the topology and evolution of chromospheric magnetic fields is necessary for linking the photospheric fields to the corona.
Using the photospheric magnetic field as a lower boundary condition for magnetic field extrapolations is questionable due to the forced nature of the former. Thus, having a reliable indicator of the magnetic field at the base of the corona is key for constraining the coronal magnetic topology and relating it to its photospheric drivers. This is a crucial piece in the end-to-end understanding of space weather and the improvement its forecasting.

Non-LTE spectral line inversion codes that include as many physical ingredients as are necessary to interpret the spectral line radiation are going to be far superior at obtaining physically meaningful and accurate results. Not only they are able to extract magnetic field information from the polarization spectra, but they also shed light on the thermodynamic stratification of the chromosphere \citep[][just to mention a few]{socas-navarro, pietarila, jaime_inversions, jaime_mg}, and even detect the presence of discontinuities \citep[see, for instance][]{alberto_calcio}. However, theory and methods are still under development and mostly used in an exploratory manner. 
At this point in time, some of the most challenging aspects of non-LTE inversion codes are their high computational cost, their finicky initialization procedures, and their unstable nature when it comes to automating them for large datasets. Therefore, they are not yet suitable for real-time quick-look data analysis or qualitative analysis of large databases. Milne-Eddington and constant slab inversions \citep[see][]{hazel}, albeit faster and more stable, are only applicable to a small subset of observations \citep[e.g.][]{casini, lopezariste, centeno_spicules, orozco}.

\noindent For certain applications, the WFA is currently the only viable tool to produce chromospheric magnetic field data products in a near-real time fashion. One example of this is the pipeline of the Synoptic Optical Long-term Investigations of the Sun (SOLIS) Vector Spectromagnetograph (VSM) instrument, which produces Level 2 vector magnetic field data products\footnote{SOLIS/VSM Ca {\sc ii} 8542 \AA\ Quick Look vector-magnetic field data can be found at https://nispdata.nso.edu/webProdDesc2/selector.php under {\em SOLIS/VSM 8542V QL full disk images}} for its full Stokes measurements of Ca {\sc ii} 8542 \AA. 

\noindent A qualitative analysis of the validity of the WFA used on the Ca {\sc ii} 8542 \AA\ line can be found in \cite{hammar}, who concludes that the approximation works well in general, allowing for a rapid and efficient inference of the chromospheric magnetic field vector.

The aim of this paper is to quantitatively assess the range of validity of the WFA for inferring the magnetic field strength and orientation in the chromosphere from the Ca {\sc ii} 8542 \AA\ spectral line under different spectral resolution and noise conditions. This particular line was chosen because it is currently the most commonly used line to probe the chromospheric magnetism, and several studies have deemed it one of the most promising diagnostics \citep{quinteronoda, quinteronoda2, lagg_review} given the accessibility to ground-based telescopes in the near-IR part of the solar spectrum, and the relative ease of its interpretation -- it is safe to assume complete re-distribution \citep{uitenbroek1989}, non-equilibrium ionization has negligible effects at the height of formation of the line \citep{wedemeyer}, and 1-D radiative transfer computations seem to deliver results that are useful in many scenarios \citep{jaime_model}.
Section \ref{sec:wfa} describes the WFA in detail and the methodology used in this analysis. Then, the WFA is applied to synthetic spectra emerging from different magnetic scenarios in Section \ref{sec:method}. A study of the effects of noise is conducted in Section \ref{sec:noise}, followed by a discussion and conclusions.

\section{The Weak Field Approximation}\label{sec:wfa}

When the Zeeman splitting ($\Delta\lambda_{\rm B}$) of a spectral line is much smaller than its typical width ($\Delta \lambda_{\rm D}$), one can apply a perturbative scheme to the polarized radiative transfer equation and reduce it to a much simpler set of equations that are valid only in the weak magnetic field limit. The mathematical condition for this is:

\begin{equation}
\bar g \frac{\Delta\lambda_{\rm B}}{\Delta \lambda_{\rm D}} << 1 \label{wfacondition}
\end{equation}

\noindent where $\bar g$ is the effective Land\'e factor of the spectral line, and the Zeeman splitting (expressed in Angstroms):

\begin{equation}
\Delta\lambda_{\rm B} = 4.67 \times 10^{-13}\lambda_0^2 B 
\end{equation}

\noindent depends on the magnetic field strength, $B$ (expressed in gauss), and the central wavelength of the spectral line, $\lambda_0$ (expressed in \AA). The Doppler width (also in wavelength units)

\begin{equation}
\Delta\lambda_{\rm D} = \frac{\lambda_0}{c}\displaystyle \sqrt{\frac{2 K_{\rm B} T}{m} + \xi^2}
\end{equation}

\noindent increases with the temperature, $T$, and the microturbulent velocity, $\xi$, whilst it decreases with the mass of the atom, $m$. Here, $K_{\rm B}$ is the Boltzmann constant.

\noindent Following the derivation in \cite{landi_book}, one can arrive at first and second perturbative order differential equations that relate the circular and linear polarization profiles to the first and second derivatives of the intensity with respect to  wavelength, respectively. These relationships provide a means of calculating the components of the vector magnetic field under a set of more or less restrictive assumptions about the variation of certain physical parameters along the line of sight.

The first order perturbation yields:

\begin{equation}
V = -\Delta\lambda_B f \bar{g} \cos\theta \frac{\partial I}{\partial \lambda}\label{eq:wfablos}
\end{equation} 

\noindent implying that Stokes $V$ is proportional to the first derivative of the intensity profile with respect to wavelength, with a proportionality factor that depends on the longitudinal component of the magnetic field, $B_{\rm LOS} = B \cos \theta$ (here, $\theta$ is the angle between the line-of-sight and the direction of the magnetic field vector). In this expression, $f$ stands for the filling factor, i.e. the fractional area of the resolution element permeated by magnetic field.

The second order perturbation leads to expressions that connect the linear polarization with the first and second derivatives of Stokes $I$ through the transverse component of the magnetic field, $B_{\rm T}= B \sin\theta$. If we choose a reference frame for the linear polarization such that Stokes $U$ is zero, we arrive at different expressions that depend on the wavelength range in which they are applied:

\begin{equation}
Q(\lambda_0) = -\frac{1}{4} \Delta\lambda_{\rm B}^2 f \bar G \sin^2\theta \displaystyle \frac{\partial^2 I}{\partial\lambda^2} \qquad \text{for } \lambda = \lambda_0 \label{eq:wfaQcenter}
\end{equation}

\begin{equation}
Q(\lambda_{\rm w}) = \frac{3}{4} \Delta\lambda_{\rm B}^2 f \bar G \sin^2\theta \displaystyle \frac{1}{\lambda_{\rm w} - \lambda_0}\displaystyle \frac{\partial I}{\partial\lambda} \qquad \text{for }  \lambda_{\rm w} \text{ in the wing of the line} \label{eq:wfaQwing}
\end{equation}

\noindent where $\bar G$ is the Land\'e factor for the transverse magnetic field, which can be obtained from combinations of the second-order moments of the Zeeman components, and is specific to the energy levels involved in the transition. Eq. \ref{eq:wfablos} requires that $B_{\rm LOS}$ is constant along the line of sight, whilst Eqs. \ref{eq:wfaQcenter} and \ref{eq:wfaQwing} assume that the plasma velocity, the transverse component of the magnetic field and its azimuth in the transverse plane, are all constant along the line of sight.  

\noindent The azimuth of the magnetic field in the plane of the sky (POS), $\chi$, can be computed using:

\begin{equation}
\frac{U(\lambda)}{Q(\lambda)} = \tan 2\chi \qquad \forall \lambda
\label{eq:azimuthwfa}
\end{equation}

\noindent which is valid for all wavelengths provided $\chi$ is constant along the line of sight.
In what follows, $Q >0$ is the reference direction for the azimuth, which increases counterclockwise in the POS. 

\subsection{A note on the magnetic filling factor}

 The magnetic filling factor is assumed to be $1$ throughout this manuscript. We will be working under the idealized assumption that the spectropolarimetric profiles emerge from an atmosphere with a spatially uniform magnetic field, although gradients along the line of sight will be considered. 

\noindent The WFA does not provide a means for determining $f$ independently of the components of the magnetic field vector. When the magnetic field is unresolved in an observation, and a fraction of the pixel is non-magnetized, all it allows us to recover is $\phi=f B_{\rm LOS}$ and $f B_{\rm T}^2$. The quantity $\phi$ is a measure of the average magnetic flux density in the resolution element, and is often (erroneously) referred to as magnetic flux. On the other hand, the quantity $f B_{\rm T}^2$, referred to in the literature as horizontal magnetic flux, measures a vector magnitude across a surface that is in general not perpendicular to the direction of said vector, which by definition is not a flux. $B_{\rm LOS}$ and $B_{\rm T}$ cannot be recovered directly, posing a severe limitation that stifles the retrieval of the intrinsic field strength, which in turn precludes the calculation of the magnetic energy density.

One can call out reasons why it may be acceptable to assume a filling factor of $f=1$, namely that the fields in the chromosphere are thought to expand laterally and fill the entire area of the resolution element (provided this resolution element is small enough). However, as the (anonymous) reviewer of this manuscript thoughtfully pointed out, there are observations that indicate that this is not the case everywhere on the Sun \citep[see][]{alberto_calcio}. The reader is advised to understand the results presented here in the context of this limitation.

\section{Methodology}\label{sec:method}
\begin{figure}
\includegraphics[angle=0,scale=.49]{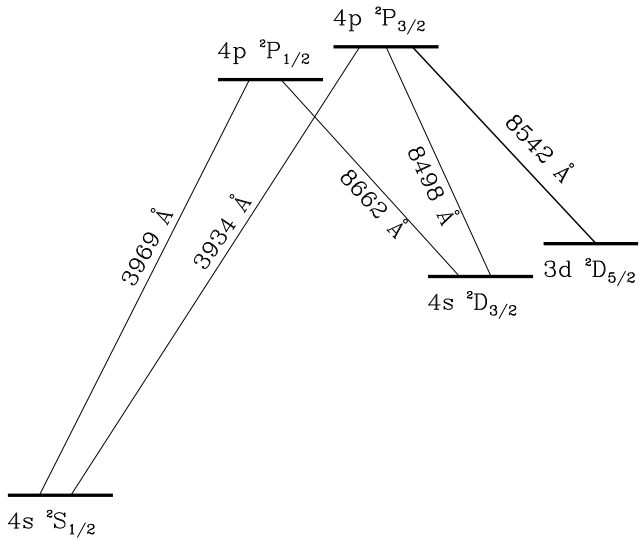}
\caption{Grotrian diagram of the Ca {\sc ii} model atom used in this work.  \label{fig:grotrian}}
\end{figure}

The validity of the weak field approximation for the chromospheric Ca {\sc ii} 8542 \AA\ is assessed on synthetic spectra created with the code NICOLE \citep{nicole, nicole2}. NICOLE is a spectral line synthesis and inversion code that solves the non-Local Thermodynamical Equilibrium (NLTE) radiative transfer equation for polarized light. It calculates the energy level populations assuming statistical equilibrium, which requires an instantaneous balance of the number of transitions to and from a given energy level; thus, no time-dependence effects can be accounted for. NICOLE also computes these populations neglecting the effects of the magnetic field, and operates in the complete frequency redistribution (CRD) regime, which is thought to be a good approximation for the Ca infrared triplet \citep{uitenbroek1989}. The spectral line polarization is induced exclusively by the Zeeman effect (polarization produced by scattering events is not considered\footnote{See section \ref{sec:scatt_pol} for a brief analysis of the effects of scattering polarization on WFA inferences}) and hydrostatic equilibrium is always imposed when the code is used in inversion mode.

\noindent The atomic model of the Ca {\sc ii} atom used in these calculations includes five bound levels as well as the Ca {\sc iii} continuum. Ca {\sc ii} 8542 \AA\ is a magnetically sensitive line that arises from the transition between the $3^2{\rm D}_{5/2}$ and the $4^2{\rm P}_{3/2}$ levels (see Grotrian diagram of Fig. \ref{fig:grotrian}). Its effective Land\'e factors can be calculated in the {\em L-S} coupling scheme and yield $\bar g = 1.1$ for the longitudinal magnetic field and $\bar G=1.18$ for the transverse field.

\noindent The 1-dimensional semi-empirical model atmosphere FAL-C \citep{falc} is used as the base model to generate the synthetic spectra. Magnetic fields of different strengths, orientations and even gradients, are introduced ad-hoc in the model in order to produce Zeeman-induced polarization signatures.

All the syntheses were carried out at disk center (for a heliocentric angle of $\mu = 1$). The spectral line was computed over a 4\AA\ wavelength range (2 \AA\ at each side of the line center), on a $5$ m\AA\ sampling grid. The synthetic spectral profiles were convolved with gaussian profiles of different widths to simulate instrument spectral smearing of varying magnitude. 
The resulting synthetic Stokes profiles were then analyzed using the weak field approximation and the results were compared to the magnetic field of the model in order to quantify the differences and qualitatively analyze the biases.
 Lastly, the exercise was repeated after adding noise of different magnitudes to the synthetic profiles, and the uncertainties of the WFA inferences were evaluated for different signal-to-noise levels.

\subsection{Constant vertical magnetic field} \label{sec:constantb}

The range of validity of the WFA for the line of sight magnetic field was assessed first. 
In this test, a number of ad-hoc constant vertical magnetic fields were added to a FALC semiempirical model atmosphere, varying the field strength between 50 and 2000 G.
The spectral line syntheses were run for several values of the macroturbulent velocity (0.5, 1, 2, 3 and 5 kms$^{-1}$) simulating different spectral resolutions and/or large-scale turbulent motions that smear the details of the spectral line to different extents.

\noindent Following \cite{marianluis}, if we perform a linear least-squares fit of the two sides of Eq. \ref{eq:wfablos}, we can extract $B_{{\rm LOS}}$ from the Stokes $I$ and Stokes $V$ spectra:

\begin{equation}
B_{\rm LOS}=-\displaystyle\frac{\displaystyle\sum_{\lambda}\frac{\partial I(\lambda)}{\partial\lambda}V(\lambda)}{C \displaystyle\sum_{\lambda}\Big(\displaystyle\frac{\partial I(\lambda)}{\partial\lambda}\Big)^2}
\label{eq:wfablosfit}
\end{equation} 

\noindent Where $C=4.66\cdot10^{-13}\bar g \lambda_0^2$ encodes information about the magnetic sensitivity of the line.
The linear regression was applied in two different wavelength ranges, yielding two sets of results: the first range used only the core of the line up to its inflection points ($\lambda_0 \pm 0.25$ \AA) and the second one used the entire synthetic spectral range ($\lambda_0 \pm 2$ \AA). The reason for this classification will become clear in Section \ref{sec:gradientb}.

\begin{figure}
\includegraphics[angle=0,scale=.49]{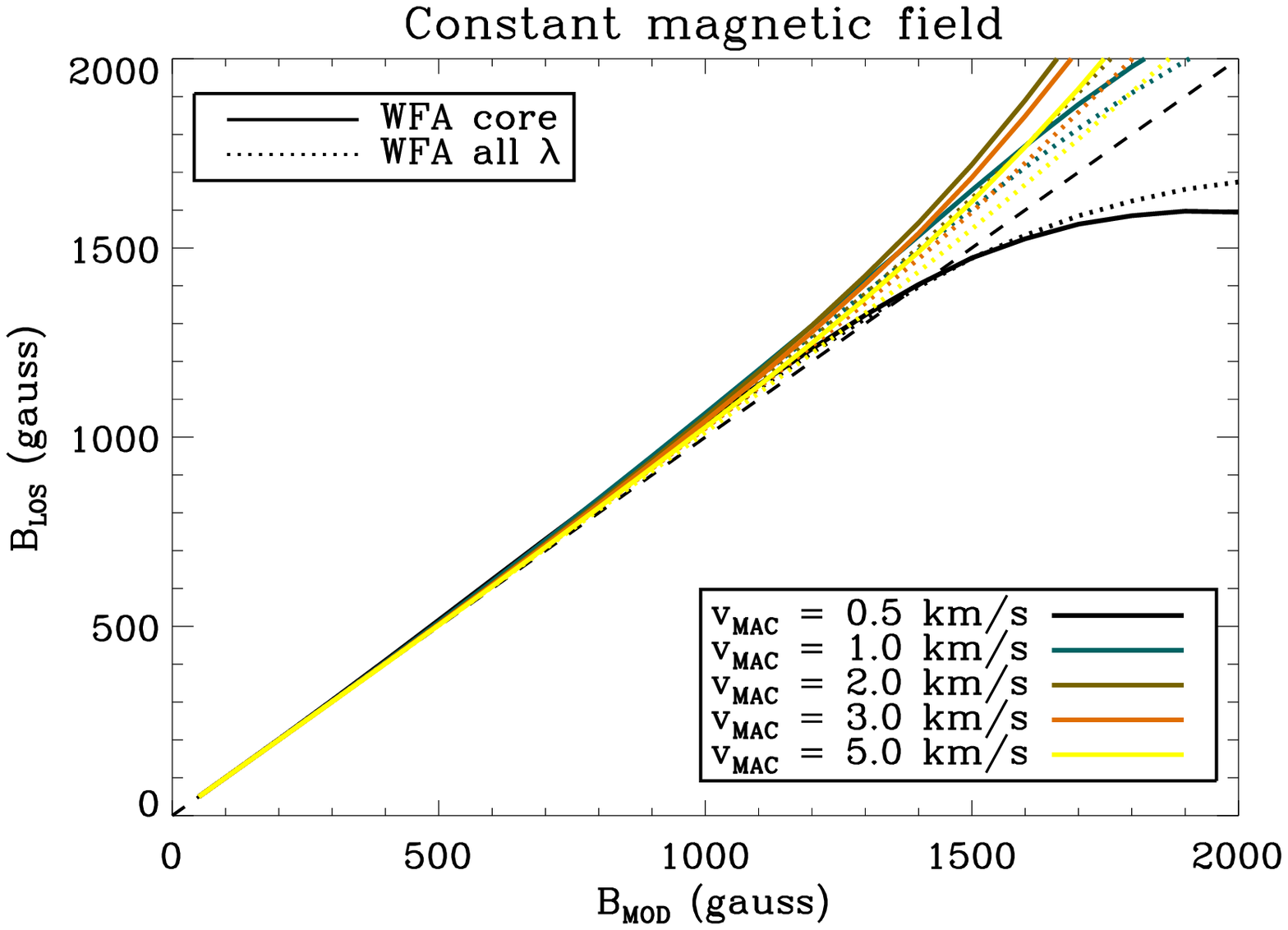}
\includegraphics[angle=0,scale=.49]{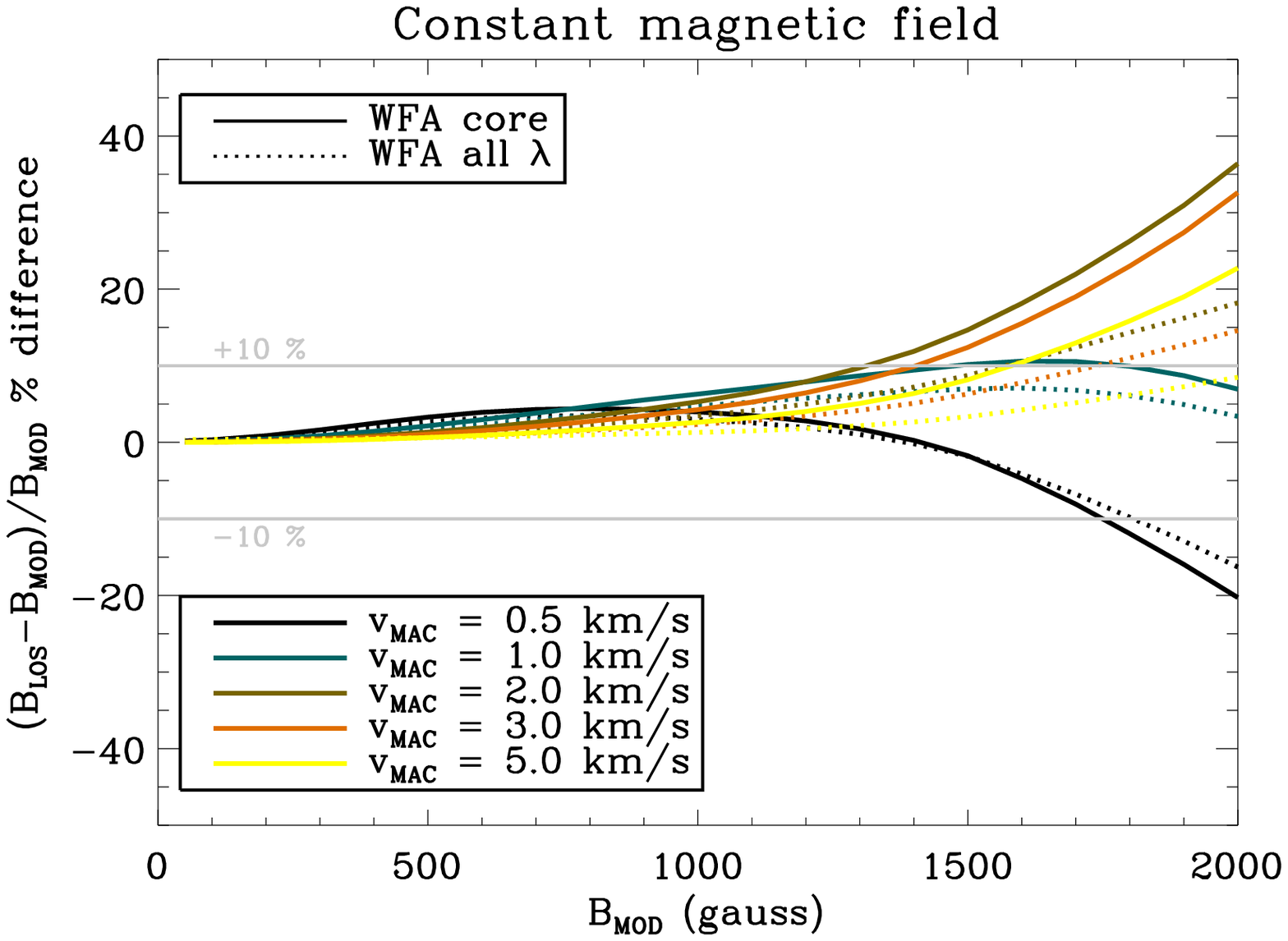}
\caption{Left: WFA inferred magnetic field strength plotted against the model value. Colors represent the different macroturbulent velocity values used in the synthesis. The solid line shows the field strengths retrieved from the core of the line whilst the dotted lines are obtain using the full 4\AA\ wavelength range. The dashed line represents the ideal one-to-one correspondence. Right: Percentage difference between the WFA inference and the model value as a function of the field strength in the model. Colors again represent macroturbulent velocity values, and dotted and solid lines refer to the full wavelength and the spectral line core inferences, respectively.  \label{fig:constantb}}
\end{figure}

\noindent Figure \ref{fig:constantb} compares the magnetic field of the model to the values inferred from the spectra using the WFA. The panel on the left shows the results using the two wavelength ranges and for all the different values of the macroturbulent velocity. The panel on the right shows the percentage difference between the WFA inference and the model value as a function of the magnetic field strength in the model. The weak field approximation in the core of the line produces results that are accurate within 10\% for magnetic fields below 1200 G. A similar accuracy is reached all the way up to 1600 G when the full spectral range of 4 \AA\ is considered in the calculation. 
In general, the results also become more accurate as the spectral smearing increases. This is because as the line becomes broader, it better satisfies the the requirement of equation \ref{wfacondition}.

\subsection{Vertical magnetic field with a gradient}\label{sec:gradientb}

\begin{figure}
\includegraphics[angle=0,scale=.49]{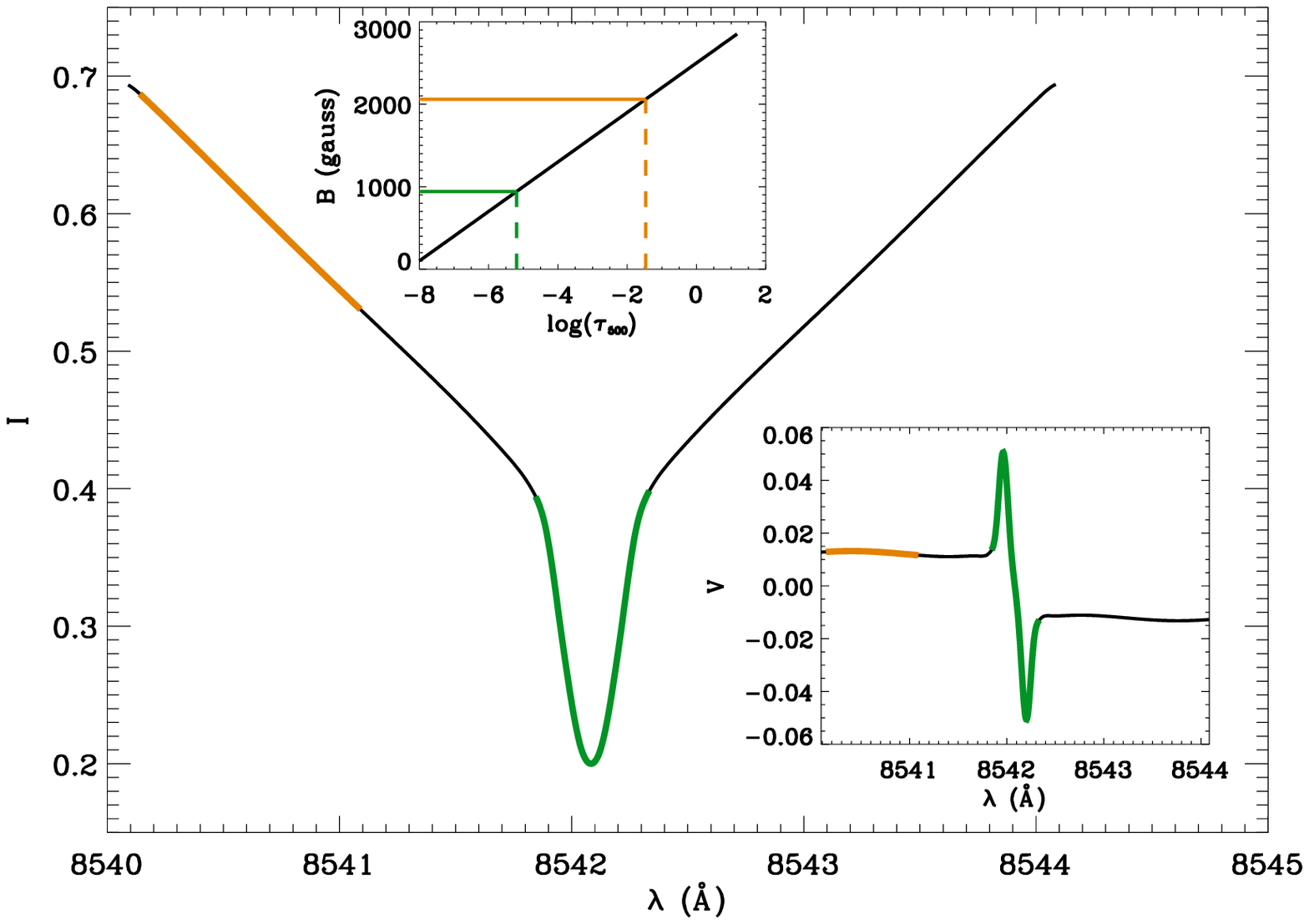}
\includegraphics[angle=0,scale=.49]{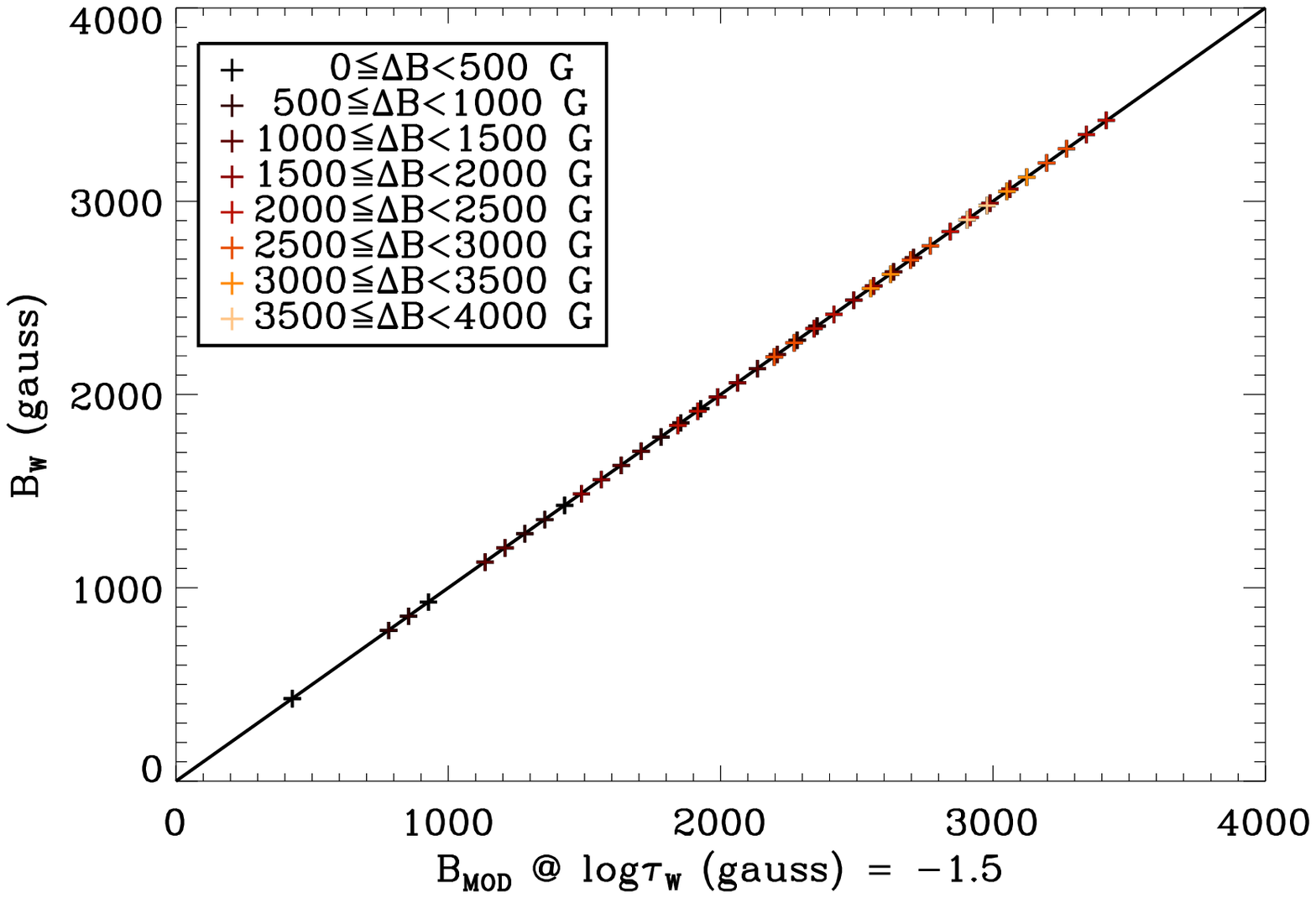} \\
\includegraphics[angle=0,scale=.49]{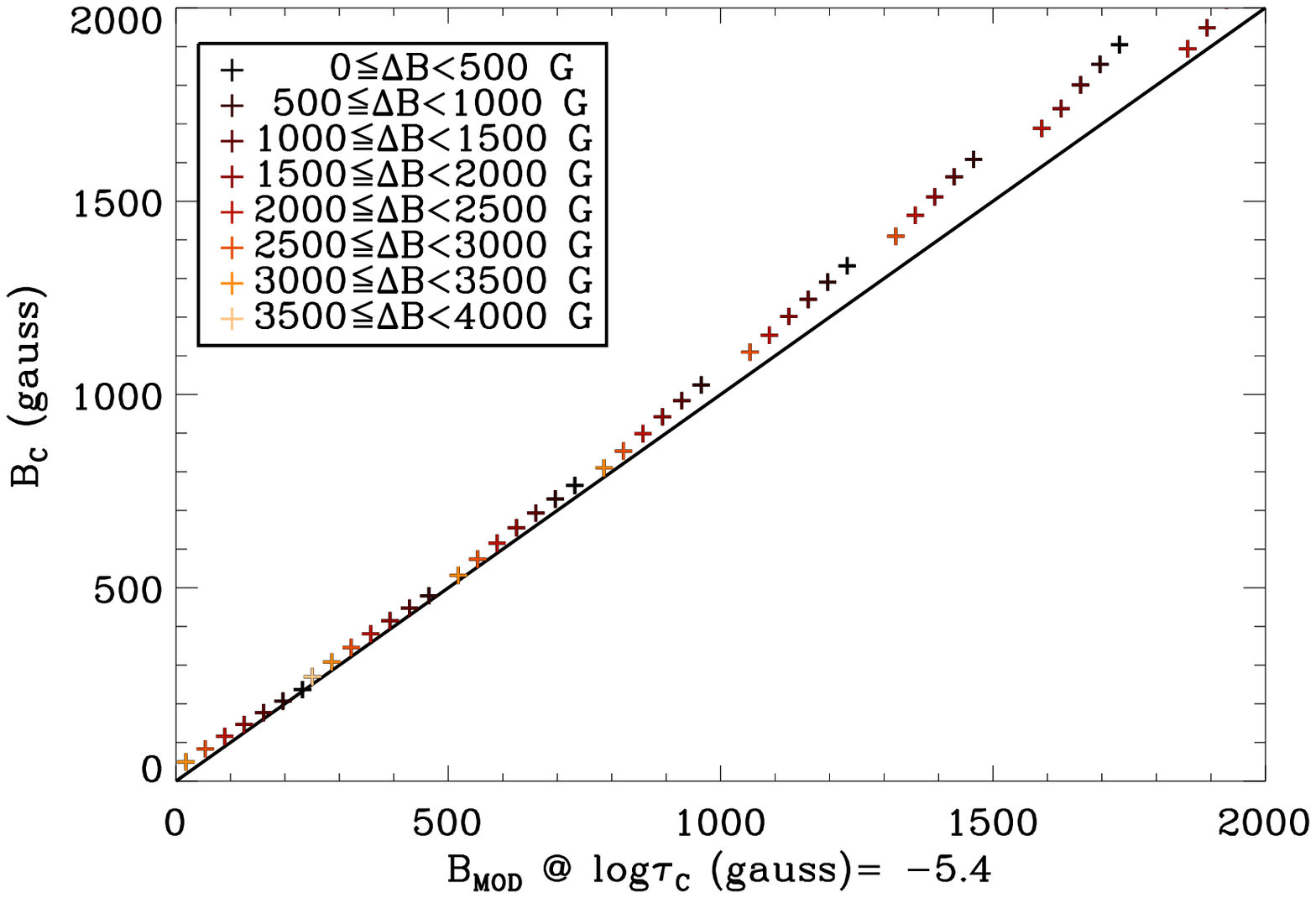}
\includegraphics[angle=0,scale=.49]{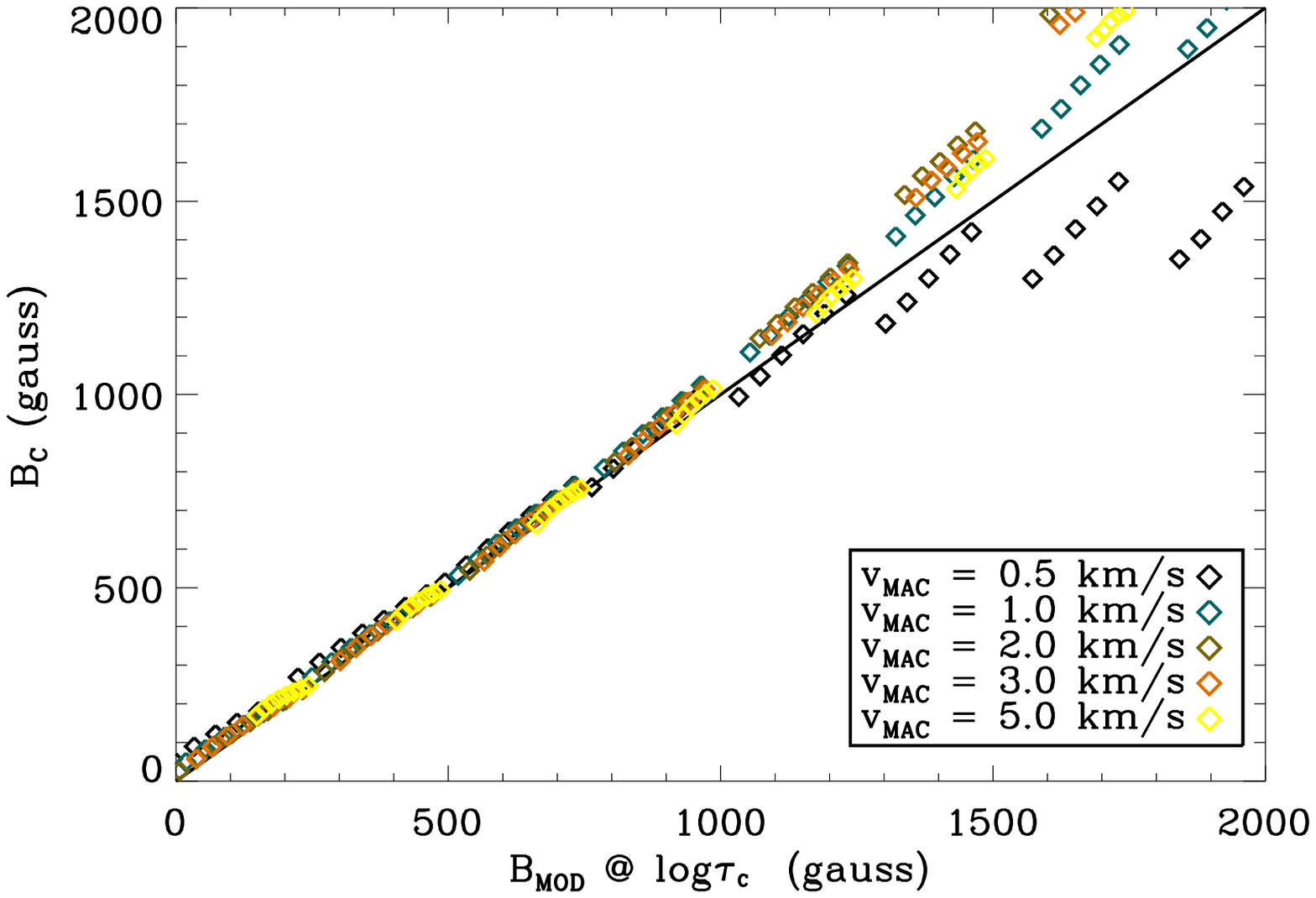}
\caption{Top Left: Synthetic Stokes $I$ as a function of wavelength. Green (orange) indicates the wavelength range used for the inference of the chromospheric (photospheric) magnetic field strength. The inset in the lower left shows Stokes $V$ using the same color convention. The inset in the upper central part of the figure shows the magnetic field of the model (in black) and the inferred magnetic field strengths for the photosphere (orange) and chromosphere (green) using the WFA, and projected (by the dashed lines) onto the corresponding height in the model. Top Right: Inferred magnetic field strength from the wing of the line versus model magnetic field at $\log\tau_{\rm w}$ . Bottom Left: Same as top right but for the line core. Bottom right: same as bottom left but for varying macroturbulent velocity values.  \label{fig:gradientb}}
\end{figure}

Magnetic fields rooted in the photosphere of the Sun tend to occupy the whole volume as they reach higher layers in the atmosphere, expanding laterally and decreasing in strength dramatically. It would be rare that a magnetic field stayed constant throughout the range of formation of the Ca {\sc ii} 8542 \AA\ line. As pointed out by \cite{quinteronoda}, the wings of this line are sensitive to a range of optical depths of $\log \tau = [0, -4]$\footnote{The reference for the optical depth in NICOLE is its value at 5000 \AA.}, whilst the line core has its maximum sensitivity higher up, in the range $\log \tau = [-4.5, -5.5]$. So the formation region of the wings and the core span more than a megameter in height altogether. 
Therefore, we decided to test the sensitivity of the WFA in the core and the wings of the spectral line separately, on a model with a non-constant vertical field. The aim was to extract information about the magnetic field strength at two different heights in the atmosphere. For this, Stokes $I$ and $V$ profiles were synthesized in FAL-C model atmospheres with ad-hoc vertical magnetic fields that varied linearly as a function of $\log\tau$. 
The depth dependence of the magnetic field of the models was generated by linearly interpolating the combinations of $B(\log\tau=0)$ and $B(\log\tau=-5)$ from the following values:
\begin{itemize}
\item $B(\log\tau=0) = [4000, 3500, 3000, 2500, 2000, 1500, 1000, 500]$ G.
\item $B(\log\tau=-5) = [2000, 1750, 1500, 1250, 1000, 750, 500, 250]$ G.
\end{itemize}
 
\noindent Only the cases for which the field strength decreased with height were computed, resulting in 48 different models with magnetic field gradients that varied from 0 to 750 G per unit of $\log \tau$.
In order to infer the magnetic field strength in the photosphere and the chromosphere, we applied the least squares fit of Eq. \ref{eq:wfablosfit} in two separate wavelength regimes for each synthetic Stokes vector: 

\begin{itemize}
\item core: $-0.25$\AA\ to $0.25$\AA\ around line center, which includes the core of the line up to its inflection points, as shown by the green part of the spectrum in the top left panel of Fig. \ref{fig:gradientb}. 
\item wing: $-2$\AA\ to $-1$\AA\ with respect to line center, in the blue wing of the Ca {\sc ii} profile, which corresponds to the orange wavelength range in the top left panel of Fig. \ref{fig:gradientb}. 
\end{itemize}

For each set of Stokes profiles we thus obtained two different values for the magnetic field strength: $B_{\rm c}$ and $B_{\rm w}$, for the core and the wing, respectively. The top left panel of Fig. \ref{fig:gradientb} shows one example of a Stokes $I$ profile (main plot) with its corresponding Stokes $V$ (inset in the lower left). The model magnetic field used to generate the Stokes profiles is represented by the black solid line in the inset at the top. In this inset, the values of $B_{\rm c}$ and $B_{\rm w}$ obtained with the WFA are shown by the solid green and orange lines. The intersection with the model magnetic field yields the average optical depths probed by the WFA in each spectral range for this particular synthetic Stokes vector.

\noindent In order to find out the atmospheric height that the inferred values of $B_{\rm w}$ are representative of, we performed a linear regression of all 48 retrieved $B_{\rm w}$ to the magnetic field in the model, $B_{\rm mod}$, treating $\log\tau$ as the free parameter of the fit (remember that $B_{\rm mod}$ is linear with $\log\tau$). This fit yielded an optical depth $\log\tau_{\rm w}=-1.4$ (in the mid-photosphere), which can be interpreted as the average optical depth for the WFA inference of the line-of-sight magnetic field obtained from the wing of the spectral line.
The top right panel of Fig. \ref{fig:gradientb} shows the $B_{\rm w}$ as a function of $B_{\rm mod}$ evaluated at $\log\tau_{\rm w}$, for the case of profiles with $v_{\rm MAC}=1$km/s. The color coding of the crosses represents the steepness of the magnetic field of the model ($\Delta B$ is the difference between the magnetic field of the model at $\log \tau = 0$ and $\log \tau = -5$). The fit is remarkably good for the entire range of magnetic field strengths and gradients tested here.

\noindent This process was repeated for the values of $B_{\rm c}$ inferred from the core of the spectral line to find the average optical depth, ${\rm log}\tau_{\rm c}$, probed by the WFA in this wavelength range. Bearing in mind the results of section \ref{sec:constantb}, the linear regression of $B_{\rm c}$ to $B_{\rm mod}$ was computed only for $B_{\rm c} < 1200$ G (because the chromospheric magnetic field strength inferred from the core of this line is typically valid in the range between 0 and 1200 G, with a ~10\% accuracy).
The bottom left panel of Fig. \ref{fig:gradientb} shows $B_{\rm c}$ as a function of $B_{\rm mod}$ evaluated at ${\rm log}\tau_{\rm c}=-5.4$ (which resulted from the afore-mentioned linear regression). Regardless of the steepness of the magnetic field gradient, $B_{\rm c}$ follows $B_{\rm mod}$ up to $\sim 1200$ G, where it starts departing significantly from the model values.

The entire exercise was repeated for a range of macroturbulent velocities. The bottom right panel of Fig. \ref{fig:gradientb} shows the behavior of $B_{\rm c}$ inferred from the line core for different values of $v_{\rm MAC}$. This panel resembles that of the constant field case (the left panel of Fig. \ref{fig:constantb}). The gradient of the magnetic field, however, creates a significant spread in the retrieved field strengths above 1200 G for all values of the spectral smearing. It must be noted that the analogous effect is not observed when the WFA is applied to the line wing, and hence the inferred photospheric magnetic fields are not significantly affected by larger values of $v_{\rm MAC}$. This is to be expected since the convolution of the spectral line with a gaussian profile will have a bigger impact on the core than on the wings of the line. 

This analysis also revealed that the larger the macroturbulence velocity value, the lower the corresponding mean optical depth for the inferred field in the chromosphere. This is a natural result of the larger spectral smearing bringing more information from the wing into the core, and thus lowering the average height sampled by the core of the line. This effect is not noticeable for the wing inferences.
Whilst the line wing probed $\log\tau_{\rm w}\sim-1.45$, the line core sampled the range $-5.40<\log\tau_{\rm c}<-5.14$.

\subsection{Line Asymmetries }\label{sec:velocity_gradient}

\begin{figure}[!t]
\includegraphics[angle=0,scale=.48]{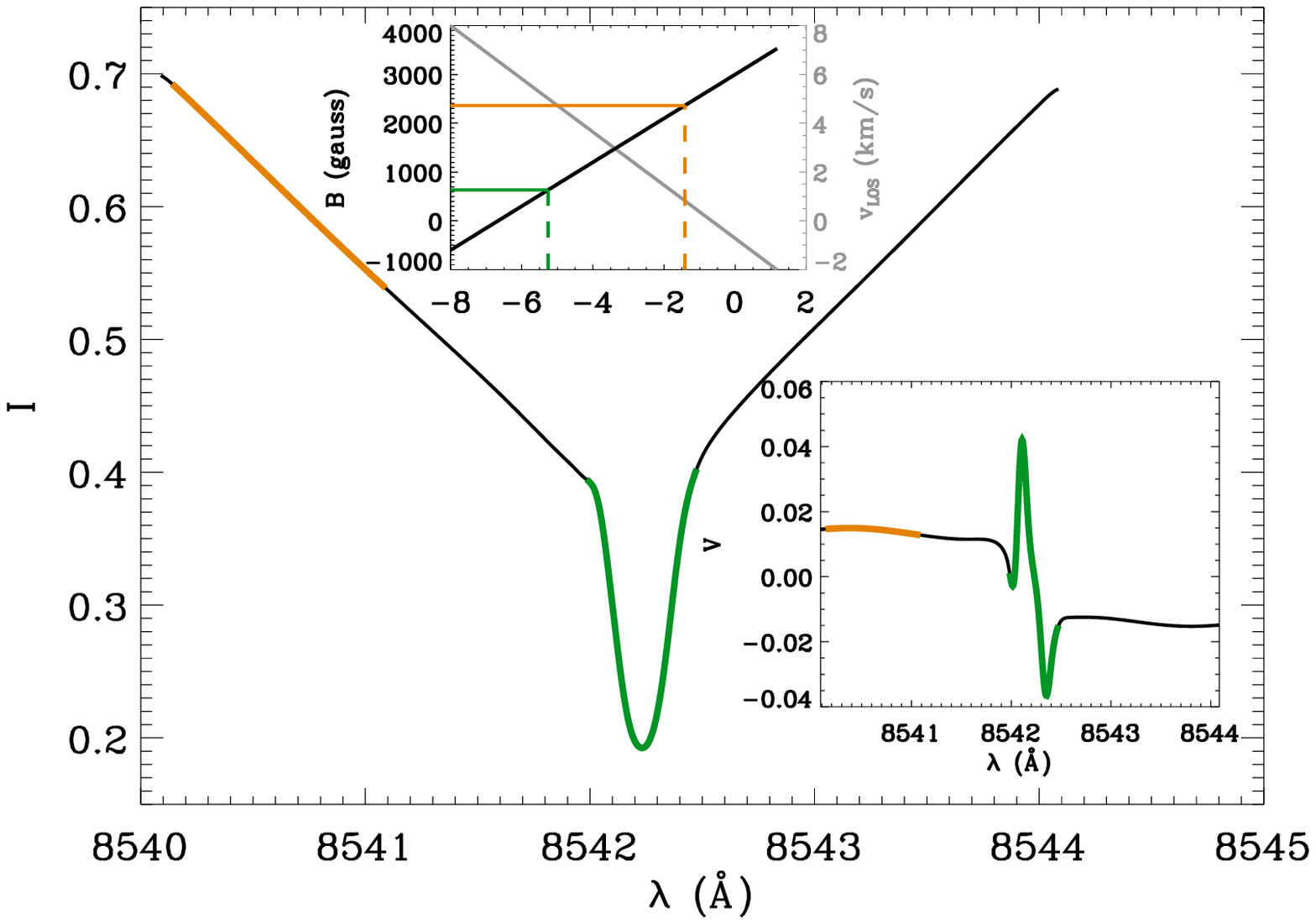}
\includegraphics[angle=0,scale=.48]{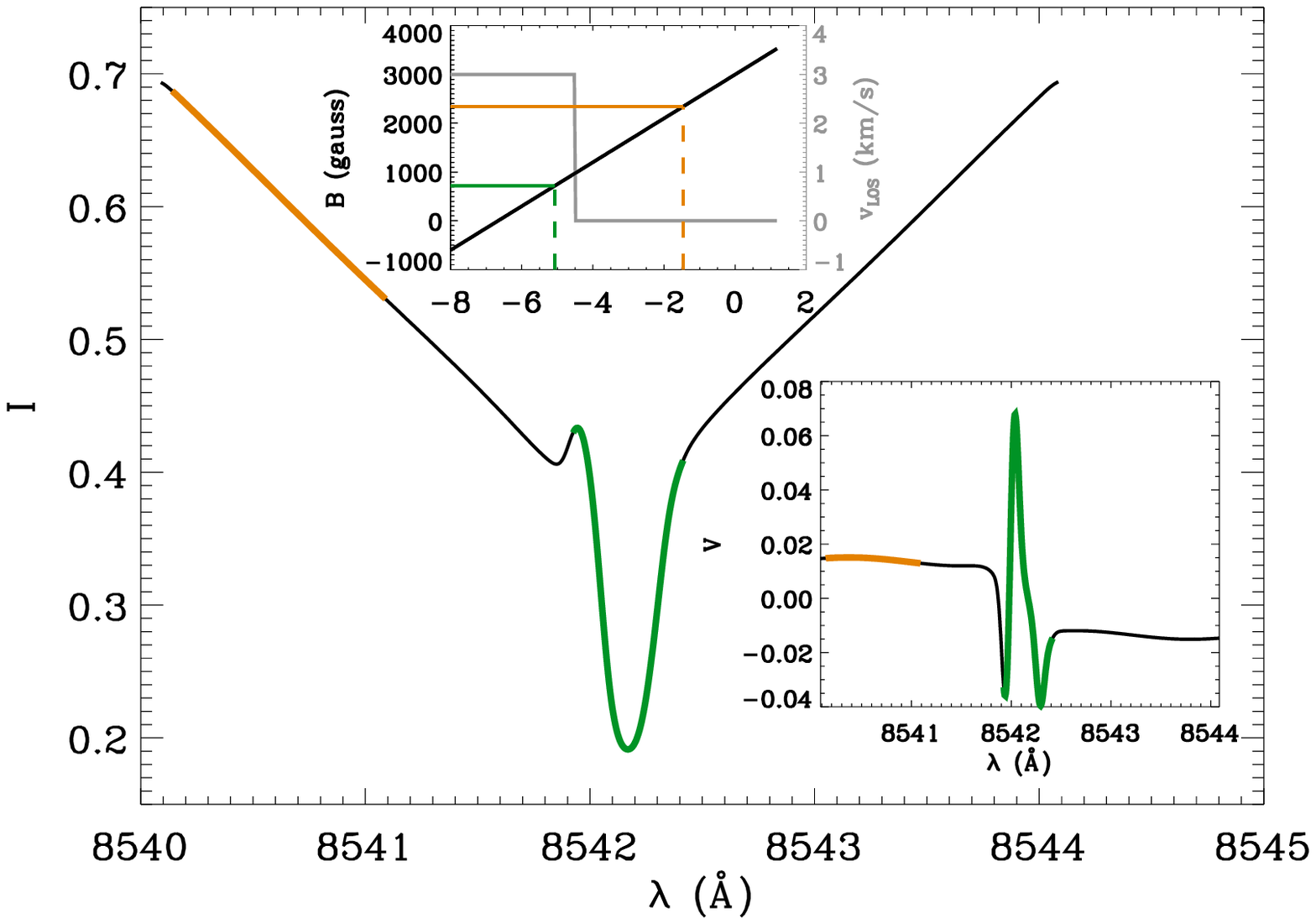}\\
\includegraphics[angle=0,scale=.48]{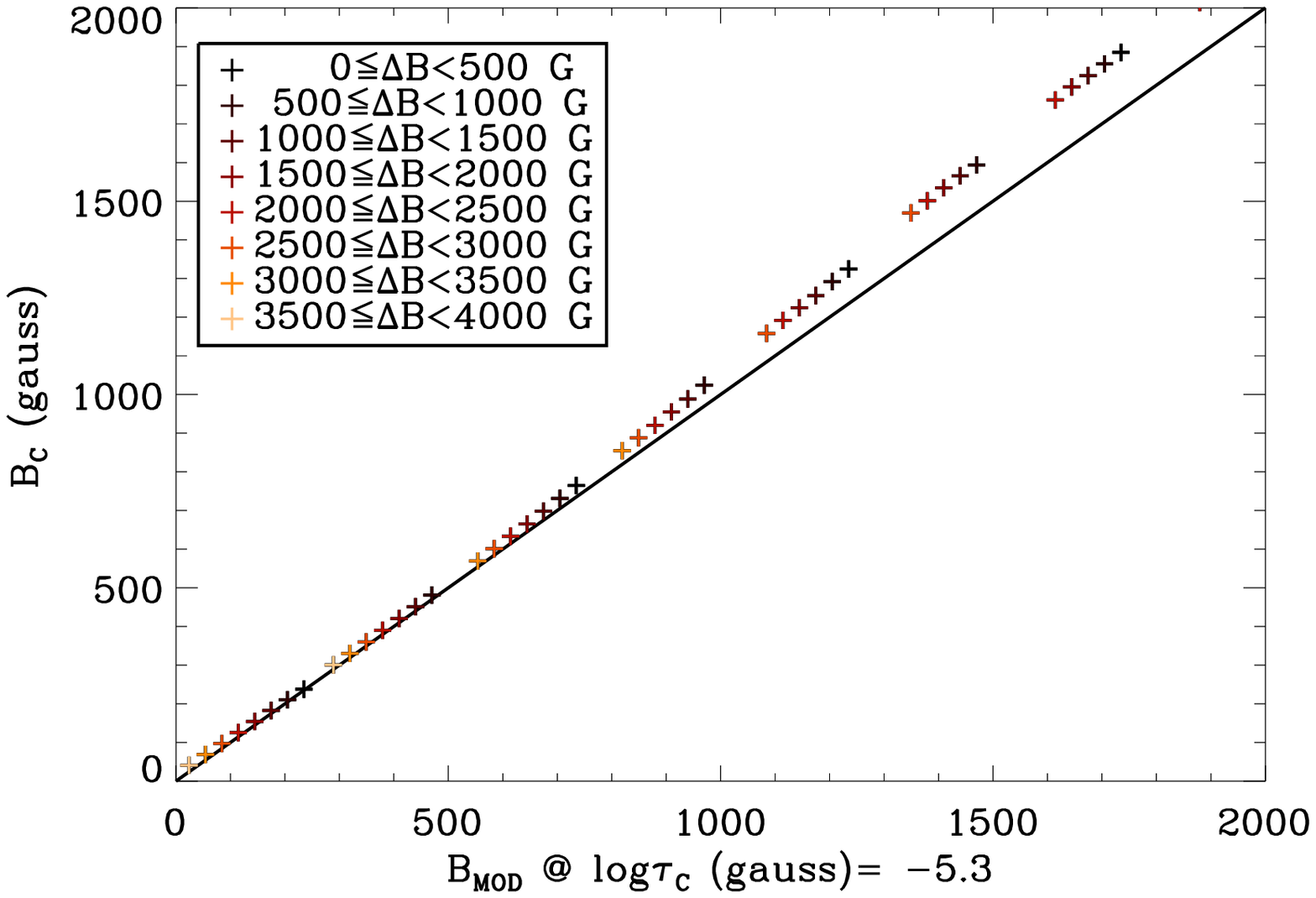}
\includegraphics[angle=0,scale=.48]{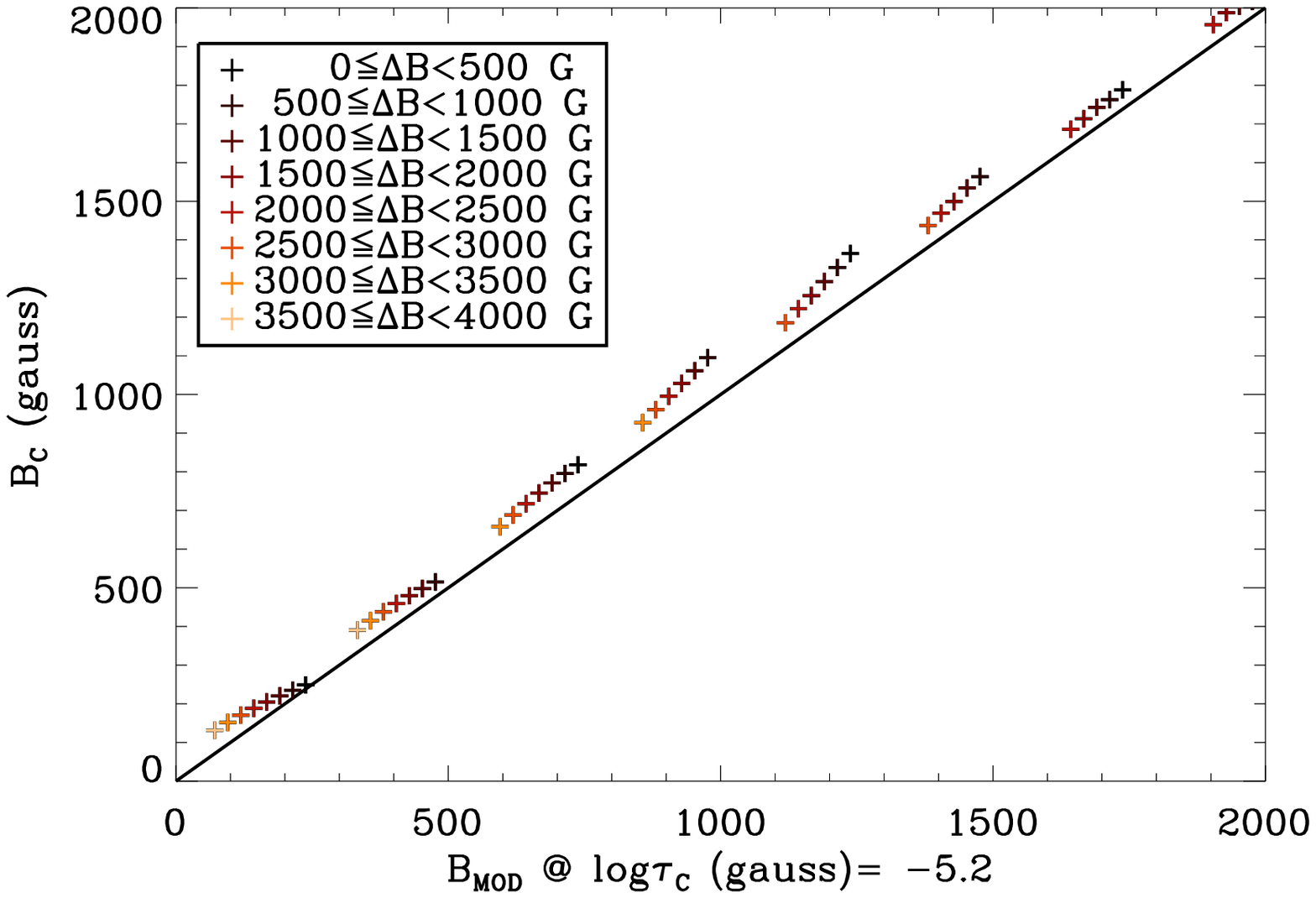}
\caption{Results from models with velocity and magnetic field gradients. The top row shows the Stokes $I$ (main plot) and $V$ (lower right inset in each panel) profiles resulting from a model atmosphere (inset in the top of each panel) with a vertical magnetic field gradient (black) and a velocity gradient (gray). The left panel shows the  case of a smooth velocity gradient while the panel on the right corresponds to a case with a velocity discontinuity at log $\tau=-4.5$. The bottom panels show how the inferred magnetic field values compare to the model atmosphere in both cases (smooth velocity gradient on the left and velocity discontinuity on the right). \label{fig:asymmetries}}
\end{figure}

 The added presence of gradients in the velocity along the line of sight is known to produce asymmetries in the Stokes spectra \cite[see, for instance][]{sanchezalmeida, pietarila}.
Both area and amplitude asymmetries between the red and blue lobes of the circular polarization \cite[see the definitions in][]{martinezpillet}, typically arise in Stokes $V$ spectra under the influence of said velocity gradients. The sight of multiple lobes and more complex spectral profiles is not uncommon either \cite[e.g.][]{viticchie}. 

In order to evaluate the effect of such asymmetries in the WFA inference, we introduced velocity gradients and discontinuities in the same model atmospheres used in Section \ref{sec:gradientb}. The resulting Stokes spectra deviate, as expected, from the canonical symmetric Stokes $I$ and antisymmetric Stokes $V$ profiles. Two different scenarios were considered.

In the first scenario, a smooth LOS velocity gradient was introduced in the model atmospheres, with a linear dependence on ${\rm log}\tau$. $v_{\rm LOS}$ changed from 0 to 5 kms$^{-1}$ between ${\rm log}\tau=0$ and ${\rm log}\tau=-5$. This corresponds to a blueshift that increases monotonically in magnitude from the the wings to the core of the Ca {\sc ii} 8542 \AA\ line.
48 different sets of Stokes profiles were generated in this velocity environment and the  various vertical magnetic field gradients of Section \ref{sec:gradientb}. The top left panel of Fig. \ref{fig:asymmetries} shows one example of the emergent Stokes profiles. The outcoming Stokes I (main figure) and Stokes V (inset in the lower right) present expected asymmetries in the blue lobe of the spectral line. 
The inset at the top center of the panel shows the magnetic field (black solid line), as well as the velocity (grey) as a function of height in the model. In the same fashion as Fig. \ref{fig:gradientb}, the orange and green lines represent the inferred photospheric and chromospheric magnetic field strengths as well as the height at which they intersect the model atmosphere.

\noindent In order to extract the chromospheric magnetic field from the line core, the $\pm 0.25$ \AA\ wavelength range was shifted to be centered around the location of the minimum of Stokes $I$ rather than the laboratory rest wavelength of the spectral line. The lower left panel of Fig. \ref{fig:asymmetries} presents the comparison between the WFA inferences and the chromospheric model magnetic fields, for all 48 sets of Stokes profiles, color-coded by the gradient of the magnetic field strength. The striking resemblance with the lower left panel of Fig. \ref{fig:gradientb} implies that the retrieved LOS magnetic field strengths are not significantly affected by the velocity gradient, especially in the $0-1200$ gauss range. However, there is a slight shift towards a lower height in the atmosphere (${\rm log}\tau_{\rm c}=-5.2$), presumably owing to the mixing of spectral information due to the velocity gradient.

Often times, the chromosphere develops shocks, and velocity gradients become velocity discontinuities. In a second scenario, a constant LOS velocity that jumps from 0 to 3 kms$^{-1}$ at ${\rm log}\tau=-4.5$ was introduced in the model atmospheres. This discontinuity is purposely placed in a region of sensitivity of the line core, to see the effects on the retrieved $B_{\rm c}$ values. The top right panel of Fig. \ref{fig:asymmetries} shows the resulting Stokes $I$ and $V$ profiles for one of the 48 model atmospheres (shown in the top central inset). The asymmetries  in the line profile are now more prominent than for the case of a velocity with a smooth gradient, and Stokes $V$ clearly presents a 3-lobed structure.
After following the same procedure than in the previous velocity scenario, the inferred chromospheric magnetic field strength is compared to that of the model atmospheres (lower right panel of Fig. \ref{fig:asymmetries}). The fidelity of the values retrieved through the WFA is now worse, indicating that a velocity discontinuity at the height of formation of the core of the spectral line, significantly impacts the inference of chromospheric field strengths, even in the $0 - 1200$~G range.

Lastly, it is worth mentioning that the photospheric inferences obtained from the wing of the line are not affected by these asymmetries due to moderate line-of-sight velocities.


\subsection{Constant magnetic field with a transverse component}\label{sec:horizontal}

\begin{figure}
\includegraphics[angle=0,scale=.48]{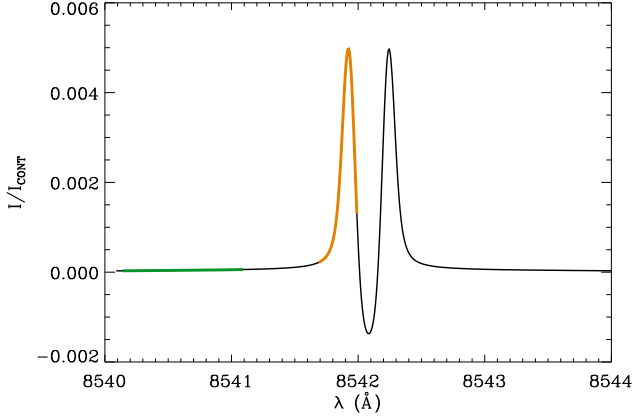}
\caption{Example of a Stokes Q profile generated for a model atmosphere with a magnetic field of 1000~G, $60^{\circ}$ of inclination and an azimuth in the transverse plane of $22.5^{\circ}$. The orange section corresponds to the wing-core boundary wavelength range used in the calculations of $B_{\rm T}$ and $\chi_{\rm WFA}$, whilst the green portion shows the wing range used for photospheric calculations. \label{fig:Qrange}}
\end{figure}

According to \citet{landi_book}, the weak field approximation for the transverse component of the magnetic field can be calculated in different ways that are valid in different wavelength ranges of the spectral line. If we modify equations \ref{eq:wfaQcenter} and \ref{eq:wfaQwing} to use the total linear polarization, $L=(Q^2+U^2)^{1/2}$, instead of Stokes $Q$:

\begin{equation}
L(\lambda_0) = -\frac{1}{4} B_{\rm T}^2 C_{\rm T} \left| \frac{\partial^2 I}{\partial \lambda^2}\right |\qquad\text{ for } \lambda = \lambda_0 \label{eq:btrans_center}
\end{equation}

\begin{equation}
L(\lambda_{\rm w}) = \frac{3}{4} B_{\rm T}^2 C_{\rm T} \left| \frac{1}{\lambda_{\rm w}-\lambda_0} \right | \left|\frac{\partial I}{\partial\lambda_{\rm w}}\right| \qquad \text{ for } \lambda = \lambda_{\rm w} \label{eq:btranswing}
\end{equation}

\noindent where $C_{\rm T}= [4.6686\cdot 10^{-10} \cdot \lambda_0^2]^2 \bar G$ is a constant for a given spectral line, $B_{\rm T}$ is the transverse component of the magnetic field, and $\bar G$ can be obtained from combinations of the second-order moments of the Zeeman components, and is specific to the energy levels involved in the transition. While Eq. \ref{eq:btrans_center} is only applicable at the line center, $\lambda_0$, Eq. \ref{eq:btranswing} is valid everywhere but in the very core ($\lambda_{\rm w}$). Both equations are arrived at under the assumption that the azimuth of the magnetic field, the strength of its transverse component, and the plasma's LOS velocity are constant with height.

The restriction of Eq. \ref{eq:btrans_center} to line center renders it non-practical for the analysis of observations with very low polarization signals. The linear polarization signatures of Ca {\sc ii} 8542 \AA\ at the line center are typically between $10^{-4}-10^{-3}$ of the continuum intensity for transverse field strengths of a few hundred gauss in the chromosphere \citep[see, for instance][]{jaime_model}.
 Besides, scattering polarization signatures (which are considered neither in the NICOLE synthesis nor in the WFA inference) would have maximum prominence at line center \citep[see][]{mansosainz} . These signatures are expected to be of the order of $<10^{-4}$ at disk center, up to $5 \times 10^{-3}$ very close to the limb ($\mu=0.1$). The scattering polarization signal decreases away from the center of the line, providing another reason to not apply the WFA inference of transverse fields to the core of the Ca {\sc ii} line. 
Thus, Eq. \ref{eq:btranswing} is better suited for retrieving $B_{\rm T}$ in the chromosphere, even though as one moves away from the line center this equation will sample lower layers in the atmosphere \citep{quinteronoda}. When choosing a wavelength range to apply  Eq.\ref{eq:btranswing}, one has to find a compromise between using a large number of wavelength samples, avoiding the bulk of the signature due to scattering polarization, and not going too far out into the wings, where polarization signatures become exclusively photospheric in origin.

 To estimate a value of $B_{\rm T}$ from a wavelength range in the spectral line, a linear least squares fit between the two sides of Eq. \ref{eq:btranswing} can be used, leaving $B_{\rm T}$ as the fitting parameter:

\begin{equation}
B_{\rm T} = \left(\frac{\displaystyle \sum_\lambda \frac{4}{3} \frac{L( \lambda)} {C_{\rm T}} \left| \frac{1}{\lambda-\lambda_0}\right| \left| \frac{\partial I}{\partial\lambda}\right|}  {\displaystyle \sum_\lambda \left| \frac{1}{\lambda-\lambda_0}\right|^2 \left| \frac{\partial I}{\partial\lambda}\right|^2}\right)^{\frac{1}{2}} \qquad \text{ for } \lambda = \lambda_{\rm w}
\label{eq:btransfit}
\end{equation}

\noindent The azimuth angle in the plane of the sky can be obtained from the ratio between the averages of $Q(\lambda)$ and $U(\lambda)$, yielding:

\begin{equation}
\chi_{\rm WFA}=\frac{1}{2} \arctan \frac{\displaystyle \sum_{\lambda} U(\lambda)}{\displaystyle\sum_{\lambda} Q(\lambda)} \qquad \forall \lambda
\label{eq:azimuthfit}
\end{equation}

\noindent Note that this last equation will not provide the quadrant in which the azimuth lies. This can be inferred with the additional information provided by the signs of Stokes Q and Stokes U.

In order to test the validity of Eqs. \ref{eq:btransfit} and \ref{eq:azimuthfit}, we applied them to synthetic Stokes profiles emerging from FAL-C model atmospheres with a variation of ad hoc constant magnetic fields of different strengths and orientations. We fixed the azimuth to $\chi=22.5^{\circ}$ and created model atmospheres with the 63 combinations of the following values of magnetic field strengths and inclinations:

\begin{itemize}
\item Field strength: B = [125, 250, 500, 750, 1000, 1250, 1500] G.
\item Inclination: $\theta= 10^{\circ}, 20^{\circ}, 30^{\circ}, 40^{\circ}, 50^{\circ}, 60^{\circ}, 70^{\circ}, 80^{\circ}, 90^{\circ}$.
\end{itemize}

\noindent In addition, the synthetic profiles were smeared with gaussian macroturbulent velocities of different widths in order to evaluate the effects of the spectral smearing on the inferred values of $B_{\rm T}$ and $\chi_{\rm WFA}$.

 \begin{figure}
\includegraphics[angle=0,scale=.48]{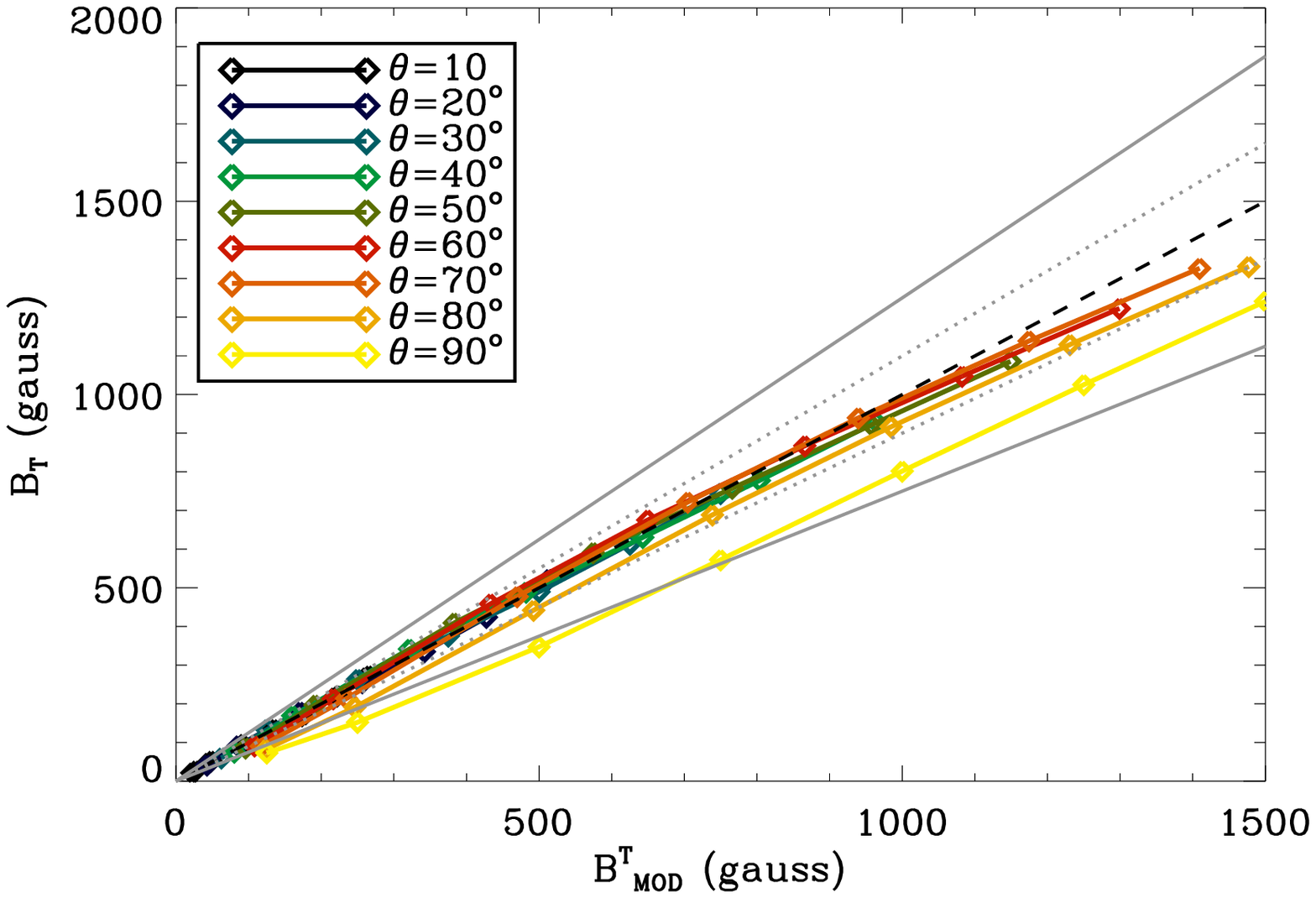}
\includegraphics[angle=0,scale=.48]{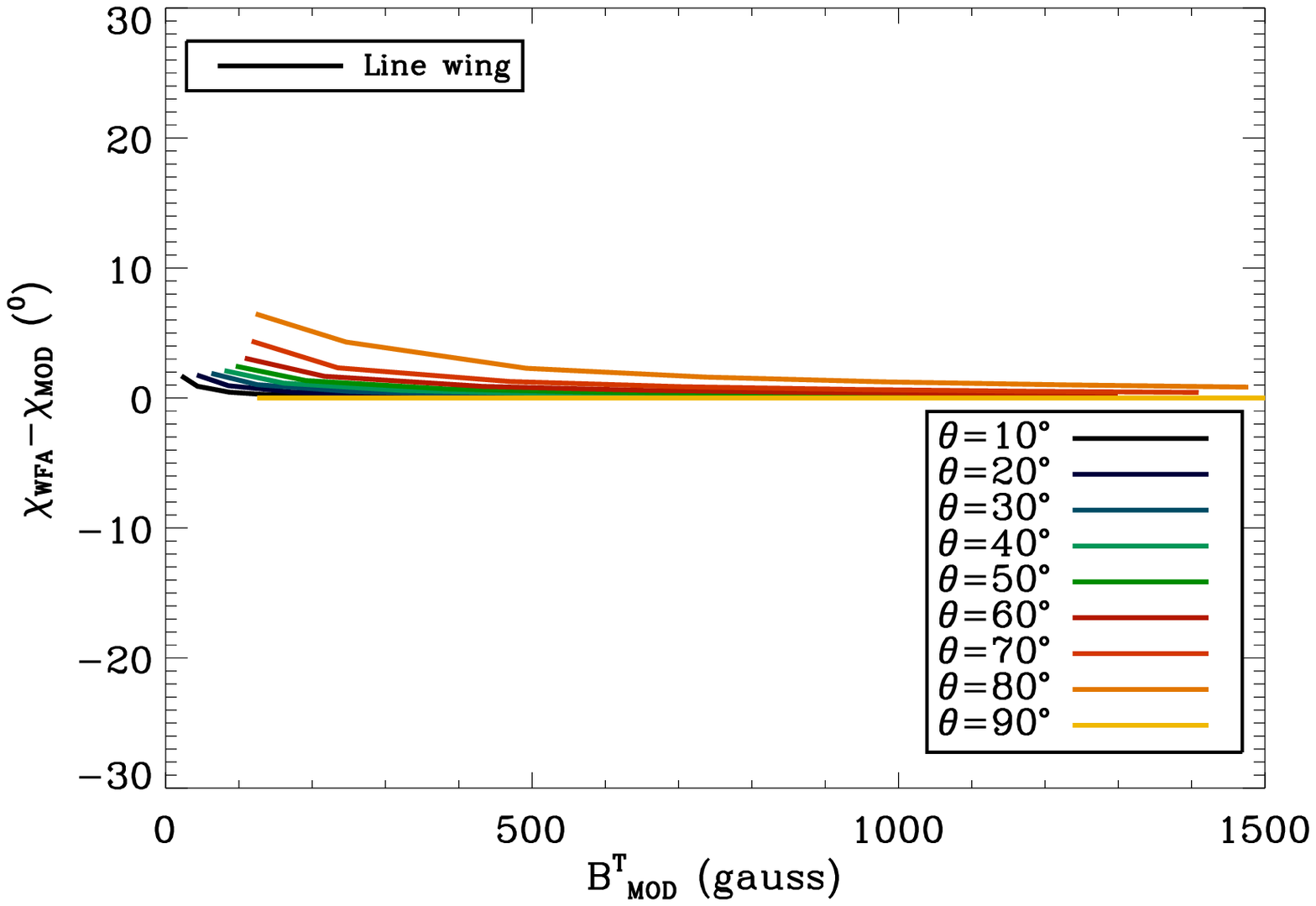} \\
\includegraphics[angle=0,scale=.48]{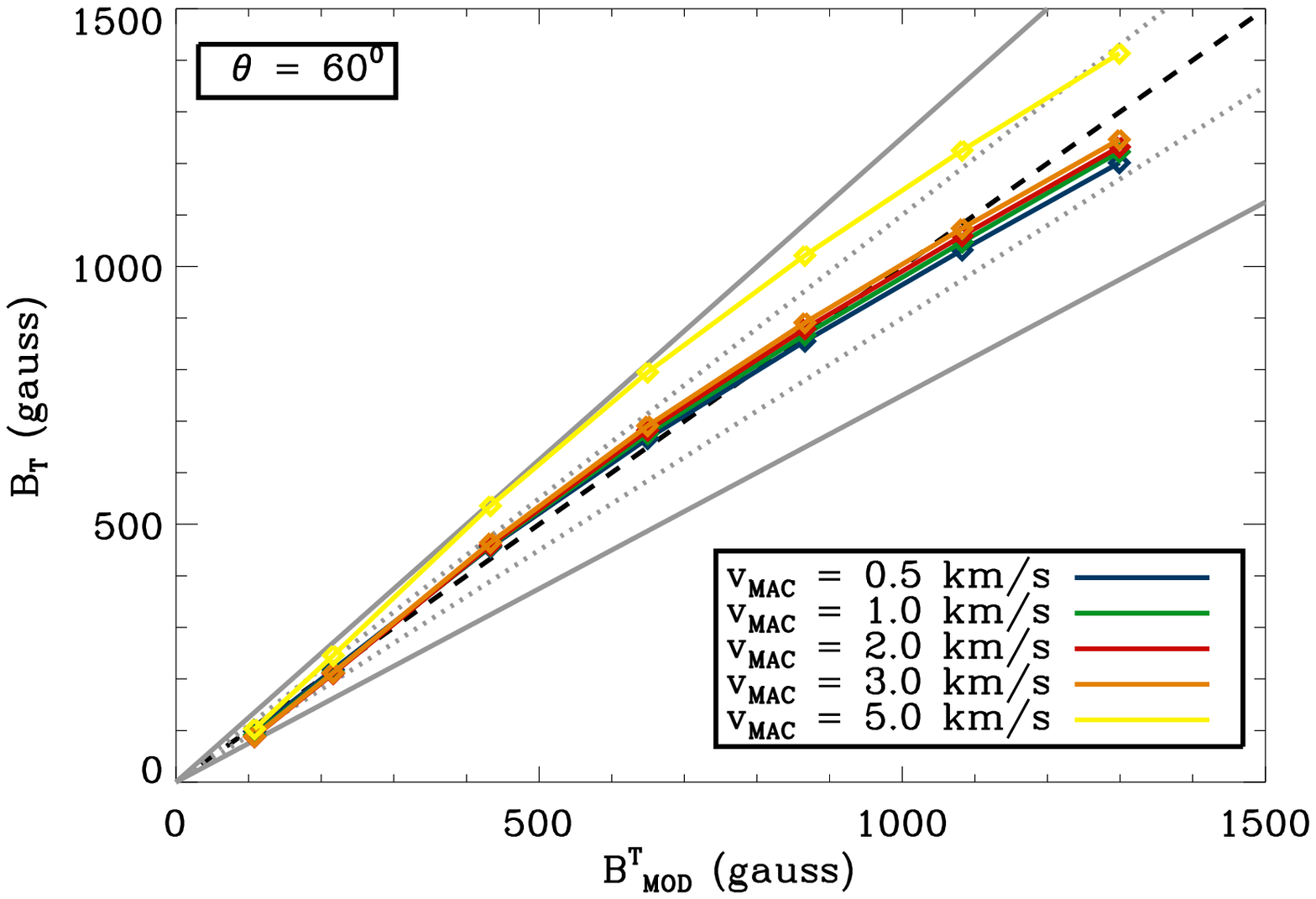}
\includegraphics[angle=0,scale=.48]{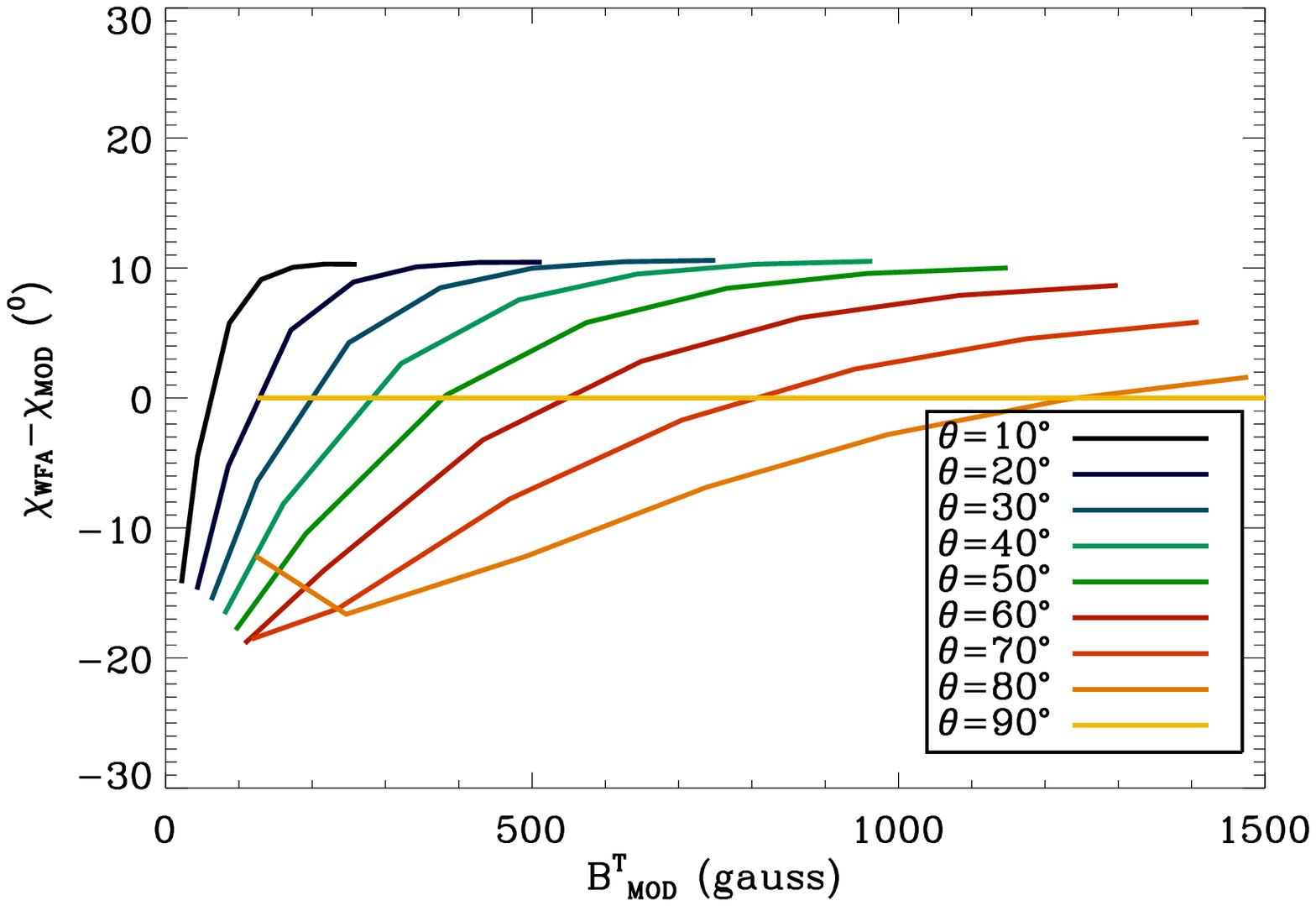}
\caption{Left: Retrieval of the transverse component of the magnetic field from the WFA as a function of the transverse magnetic field strength of the model for different inclination angles (top) and different macroturbulent velocity values (bottom). The dotted and solid gray lines represent the 10\% and 25\% accuracy boundaries, respectively. Right: Difference between $\chi_{\rm WFA}$ and model azimuth angles as a function of the transverse field strength, with Eq. \ref{eq:azimuthfit} evaluated in the wing of the line (top) and the core-wing boundary (bottom). The colors refer to the different inclination angles. \label{fig:transverseb}}
\end{figure}

As mentioned above, there needs to be a compromise in the choice of the wavelength range to apply Eq. \ref{eq:btransfit}. After multiple tests, we settled on the $\lambda-\lambda_0=[-0.4,-0.1]$, which we will refer to as the wing-core boundary (see orange sector of the Stokes Q profile shown in Fig. \ref{fig:Qrange}). This choice eliminates the line center and inner core from the computation, but does not go too far out into the wing as to incorporate significant photospheric signal \citep[see the response functions of Stokes Q and U in ][]{quinteronoda}. 

\noindent The panels on the left side of Fig. \ref{fig:transverseb} compare the transverse magnetic field strength obtained from Eq. \ref{eq:btransfit} to that of the known model input. The top panel shows $B_{\rm T}$ for the case of $v_{\rm MAC} = 1$km/s and different inclination angles, whilst the bottom panel fixes the inclination to $\theta=60^{\circ}$ and exposes the effects of varying the macroturbulent velocity. In both panels, the dashed line represents the ideal solution whilst the dotted (solid) grey lines mark the 10\% (25\%) accuracy boundary. All of the results are accurate within $25\%$ of the model value, and usually much better than that (typically within 10\%). As expected, the spread in the inferred values generally increases as the magnetic field strength of the model becomes larger, departing from the weak field assumption. However there is a significant departure from the model value in the top left panel, as the inclination angle reaches $90^{\circ}$, resulting in a severe underestimation of $B_{\rm T}$. 
In the bottom left panel, on the other hand, it becomes evident that the larger value of the macroturbulent velocity consistently yields larger values of $B_{\rm T}$, particularly for the case of $v_{\rm MAC}=5$km/s. This is presumably due to the spectral smearing, which introduces information from the inner core of the line into the computation of Eq. \ref{eq:btransfit}, which breaks down at $\lambda \sim \lambda_0$.

{ The panels on the right side of  Fig. \ref{fig:transverseb} evaluate the accuracy in the determination of the azimuth of the magnetic field for different inclination angles (colors) using Eq. \ref{eq:azimuthfit}, which was evaluated in the wing ($\lambda-\lambda_0=[-2, -1]$ \AA) and the core-wing boundary ($\lambda-\lambda_0 =[-0.4, -0.1]$ \AA) of the spectral line. This yielded estimates of the photospheric (top) and chromospheric (bottom) magnetic field azimuths. The azimuth angle of the model, $\chi_{\rm mod}$, was $22.5^{\circ}$ in all cases, so these panels represent the difference between $\chi_{\rm WFA}$ and $\chi_{\rm mod}$ as a function of the transverse field strength. 

\noindent The results are insensitive to the macroturbulent velocity of the model (not shown in the figure), but present a strong dependence with the magnetic field strength and its inclination angle. 
The absolute difference between the inferred and the model azimuths never exceeds $7^{\circ}$ when calculated in the wing of the line (top), it decreases with increasing magnetic field strength, and the error is larger for larger inclination angles. However, this wavelength range samples the photospheric magnetic field azimuth, so it is not necessarily indicative of the direction of the magnetic field in the chromosphere. Besides, the wing of the line contains very little linear polarization signal when compared to the core, rendering it even less useful for a WFA inference in the presence of noise.

\noindent The discrepancies in the azimuth retrieved from the core-wing boundary with respect to the model value (lower right panel) can be as large as $20^{\circ}$. These differences in the inferred and model azimuthal angles are consistent with what \cite{ronan} find for the Fe {\sc i} 6302.5 \AA\ line in sunspots, and are ascribed to magneto-optical effects (see Sect. \ref{sec:mo}). Despite the larger error, this wavelength range probes chromospheric layers and is thus better suited than the line wing to study the chromosphere. 
Note that for $\theta = 90^{\circ}$, the azimuth is retrieved with no error, irrespectively of the magnetic field strength and the range of wavelengths used in the WFA calculation. This is because magneto-optical (M-O) effects are minimal for purely transverse fields.

\subsubsection{Scattering Polarization}\label{sec:scatt_pol}
\begin{figure}
\includegraphics[angle=0,scale=.6]{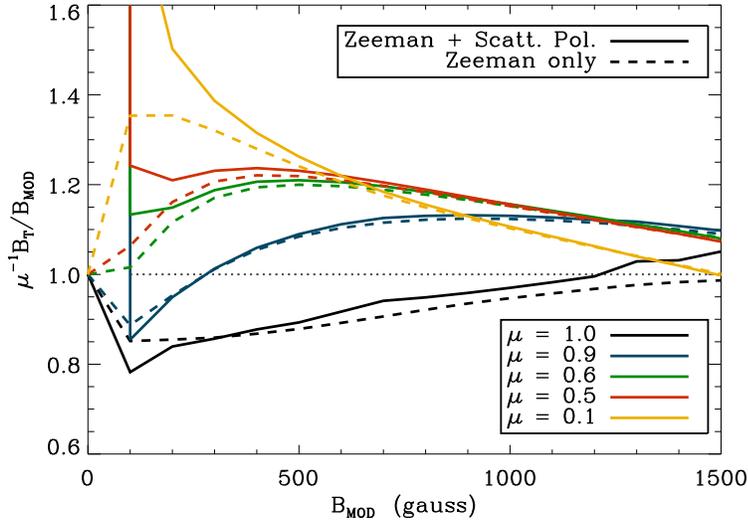}
\caption{Performance of the WFA inference on linear polarization profiles resulting from the combined action of scattering polarization, and the Zeeman and Hanle effects. This figure shows the ratio of the transverse WFA inference and the model field strength as a function of the latter. Whilst the solid lines result from the full scattering polarization calculation, the dashed lines were obtained from the ``Zeeman-only'' case. The results for different observing geometries are shown in different colors.\label{fig:scatt_pol}}
\end{figure}

In quiet Sun areas, where magnetic fields are weak, scattering polarization signatures typically dominate the linear polarization profiles of Ca {\sc ii} 8542 \AA\ \citep{mansosainz}. This is particularly true close to the limb, where when combined with the enhancing effect of shocks, the linear polarization signatures can reach amplitudes of $\sim 10^{-2}\times{\rm I_C}$ \citep{carlin}. Only in stronger field areas, the Zeeman effect will begin to dominate over the scattering polarization. 
Strong Zeeman-like linear polarization patterns are not commonly seen in the Ca {\sc ii} 8542 \AA\ line, and have only been reported in sunspot penumbrae \citep{joshi} and flaring regions \citep{kuridze}. The latter investigation presents a WFA inference of transverse field strengths of $B_{\rm T} \sim 1.3$~kG that is consistent with the results of non-LTE spectral line inversions.

A special set of calculations was carried out in order to obtain an estimate of the effect of scattering polarization on the WFA inference of the transverse magnetic field. 
Since NICOLE does not account for the generation of polarization induced by scattering processes, this set of tests was executed with the 1-D radiative transfer code for polarized complex atoms used in \cite{tanausu}. This code solves the polarized radiative transfer in a plane-parallel geometry for a polarized multi-term or multi-level atom in the presence of an arbitrary magnetic field, taking into account the effects of partial frequency redistribution, as well as the contribution of inelastic and elastic collisions. 

\noindent Two sets of calculations of the emergent Stokes profiles were carried out. One in which the linear polarization was a result of scattering phenomena as well as the combined Zeeman and Hanle effects, and one in which scattering polarization was turned off, and only polarization due to the Zeeman effect was permitted.
The emergent Stokes profiles were computed in a FAL-C model atmosphere with a constant transverse magnetic field (with strengths varying from $0$ to $1500$~G in steps of $100$~G). 
Since the amplitude of the scattering polarization signatures increases as the observing geometry approaches the limb, the emergent Stokes profiles were computed for 5 different line of sights, from $\mu=1$ (disk center) to $\mu=0.1$ (i.e. $84^{\circ}$ of heliocentric angle). 
All spectral profiles were generated under the assumption of complete frequency redistribution.

Fig. \ref{fig:scatt_pol} shows the results of the WFA inference for the transverse magnetic field from these two sets of synthetic spectral profiles. The inferred $B_{\rm T}$ (corrected by the effect of the viewing angle, $\mu$) relative to the model atmosphere value, is represented as a function of the latter. Whilst the dashed lines show the ``Zeeman-only'' case, the solid lines refer to the full calculation with scattering-induced polarization modified by the Hanle and Zeeman effects. The colors account for the different observing geometries. 

As expected, while the WFA inference from the ``Zeeman-only'' profiles converges towards the model value in the zero-field case, the opposite is true for the inferences from the scattering polarization signatures, which diverge strongly from the model at low field strengths. However, for most geometries and for field strengths above $200$~G, the discrepancies between these WFA inferences and the corresponding model values remain within 20\% of the latter. 
In the case of $\mu=0.1$ (heliocentric angle of $84^{\circ}$), this only holds true for field strengths above $600$~G.

\subsection{Magneto-Optical Effects}\label{sec:mo}

Magneto-optical effects are produced by the dispersion coefficients in the propagation matrix of the radiative transfer equation \citep{josecarlos_book}. As the light travels through the atmosphere, M-O effects introduce phase shifts between the components of the electric field vector that result in linear-to-circular polarization conversion (which is known as Faraday pulsation), or in the change of the states of linear polarization amongst themselves (this is referred to as Faraday rotation). According to \citep{landi_book}, in the weak field limit ($\Delta\lambda_{\rm B} << \Delta\lambda_{\rm D}$), M-O effects yield corrections of the order of 
$(\Delta\lambda_{\rm B}/\Delta\lambda_{\rm D})^m$, with $m=5$ for Stokes $I$, $m=2$ for Stokes $V$ and $m=1$ for Stokes $Q$ and $U$.
In order to evaluate their importance, we should estimate the magnitude of this ratio. Fig. \ref{fig:wfaratio} presents the quantity $\Delta\lambda_{\rm B} / \Delta\lambda_{\rm D}$ as a function of height in a FAL-C atmosphere with constant magnetic fields of 500G (dashed line) and 1000G (solid line). For the latter, this ratio is close to 1, implying that M-O effects will be significant in the linear polarization profiles. These effects will be much less important in Stokes $V$, owing to its quadratic dependence, and much less so in Stokes $I$, due to the power of 5 .

\begin{figure}
\includegraphics[angle=0,scale=.49]{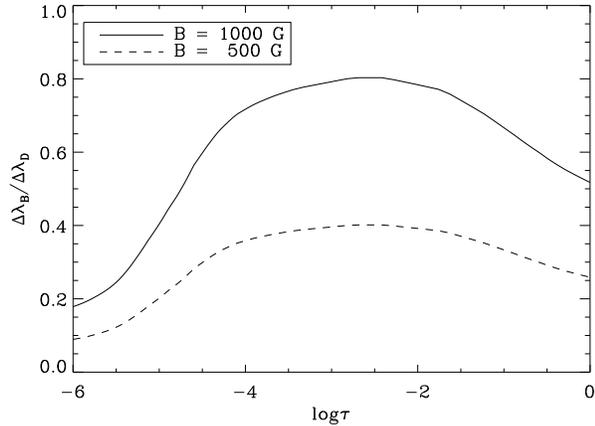}
\caption{Ratio of the Zeeman splitting and the Doppler width of the Ca {\sc ii} 8542 \AA\ line as a function of optical depth in a FAL-C atmosphere for magnetic field strengths of 500G (dashed line) and 1000G (solid line). \label{fig:wfaratio}}
\end{figure}

M-O effects introduce signatures in all Stokes profiles. In Stokes $V$ they produce a reversal feature around the line center that is sensitive to the field strength and inclination, as well as to the Zeeman pattern. However, the linear polarization signals are the most affected by M-O effects \citep{landi_book}, and can be easily seen in Kawakami diagrams, in which Stokes $U$ is plotted against Stokes $Q$, showing the ratio between the two parameters. In the absence of M-O effects, $Q$ and $U$ should display a linear relationship, with the slope of the line being equal to $\tan 2 \chi$. However, if M-O effects are present, the relationship between $Q$ and $U$ deviates from this linear trend, typically behaving better in the wings of the spectral line than close to the core.

\begin{figure}
\includegraphics[angle=0,scale=.49]{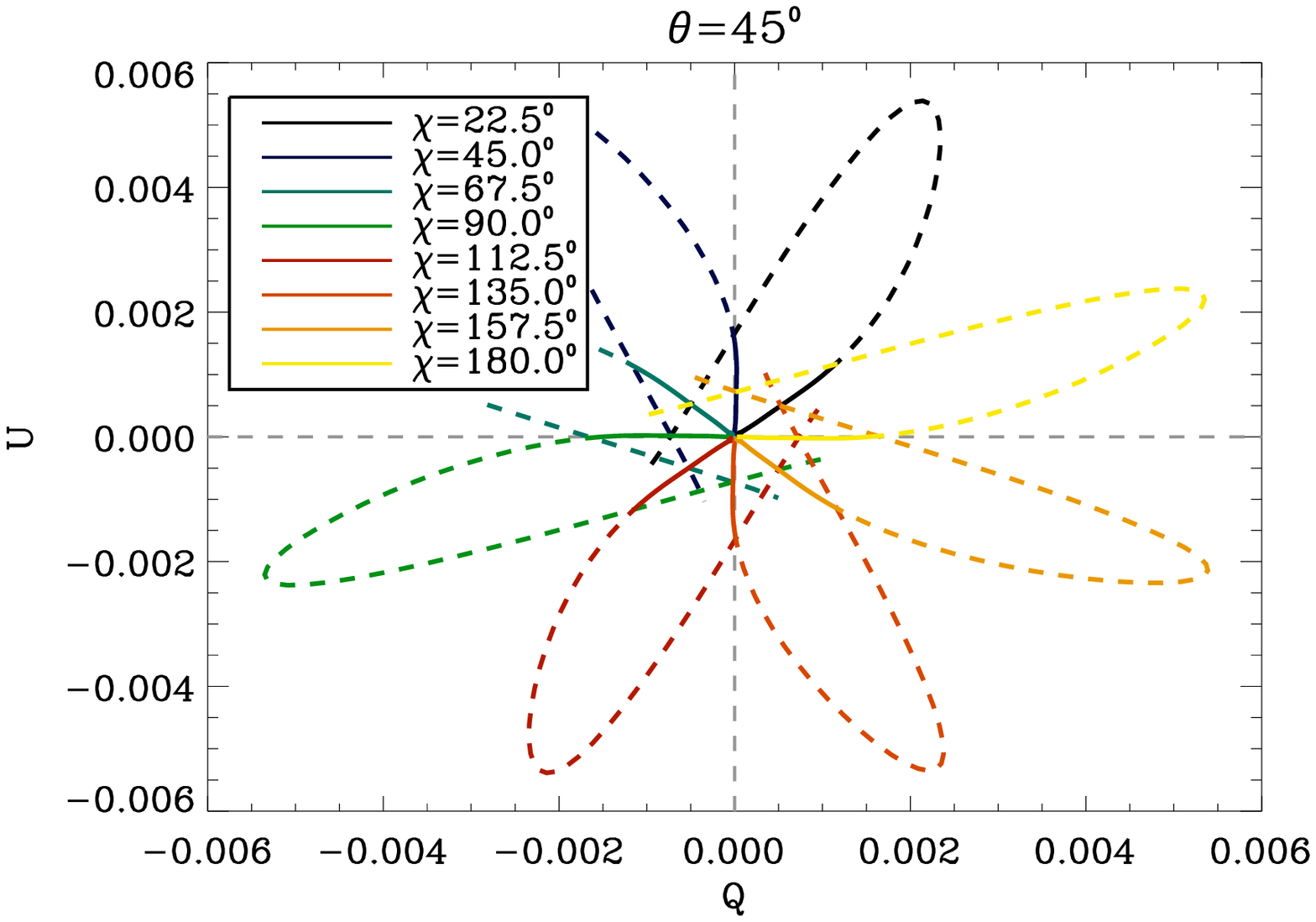}
\includegraphics[angle=0,scale=.49]{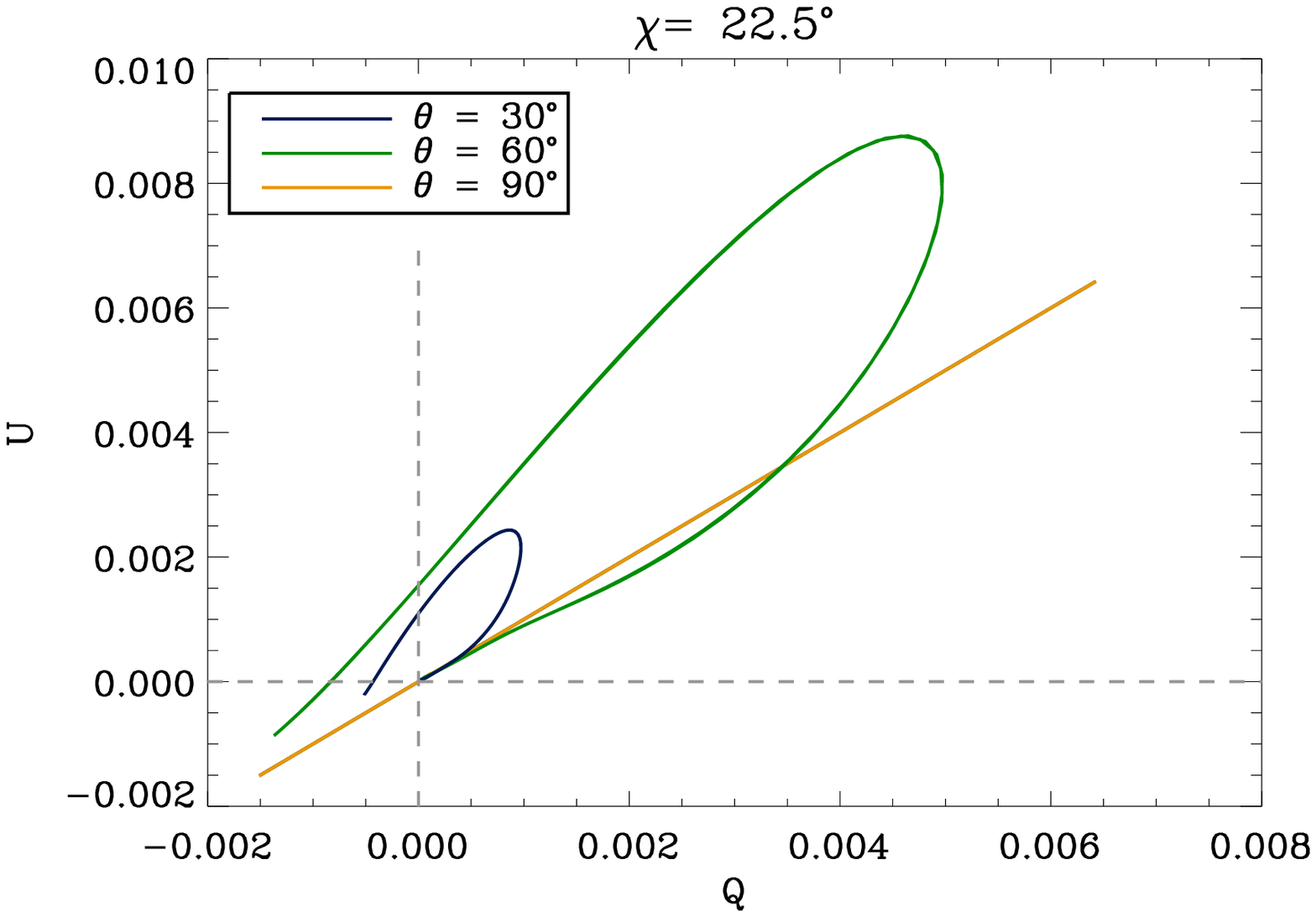}
\caption{Kawakami diagrams for a FAL-C atmosphere with a constant magnetic field of 1000G. The left panel shows the cases of various azimuth angles and a fixed inclination of $\theta=45^{\circ}$. The right panel represents the relationship between $Q$ and $U$ as the inclination takes three different values ($\theta=30^{\circ}$, $60^{\circ}$, and $90^{\circ}$) and the azimuth is fixed at $\chi=22.5^{\circ}$. \label{fig:kawakami}}
\end{figure}

Fig. \ref{fig:kawakami} shows different Kawakami diagrams for a constant magnetic field of 1000 G. The left panel represents $U$ vs. $Q$ for a magnetic field with an inclination of $\theta=45^{\circ}$ with respect to the LOS and various values of the azimuth angle in the plane of the sky. The behavior of the relationship in the wing of Ca {\sc ii} 8542 \AA\ (shown with solid lines) differs from that of the core (represented by dashed lines). In the line wing, $Q$ and $U$ present a linear behavior, with a slope that is exactly the tangent of two times the azimuth angle. In the core, however, the relationship between $Q$ and $U$ deviates from its ideal behavior due to the M-O effects, rendering the 
azimuth inference from Eq. \ref{eq:azimuthfit} incorrect. 
 
\noindent  The right panel of Fig. \ref{fig:kawakami} shows the behavior of the linear polarization when varying the inclination angle. For $\theta=90^{\circ}$ the Kawakami diagram recovers the ideal behavior, this is, a linear relationship between $Q$ and $U$. This happens because M-O effects are negligible for purely longitudinal or transverse fields \citep{landi_book}. They also disappear in the intense field limit, which could explain the better accuracy of the inferred $\chi_{\rm WFA}$ as the strength of the field increases (see the top right panel of Fig. \ref{fig:transverseb}).

\section{Effects of noise and spectral sampling}\label{sec:noise}

Real spectropolarimetric observations are, of course, always affected by noise to a varying degree. Typical noise levels for these observations are of the order of $\sigma_{\rm n}\sim 10^{-3}$ in units of the continuum intensity \citep[see, for instance,][]{ichimoto, schou, collados_gris}. In the best cases, noise has been taken down to a few times $\sigma_{\rm n} \sim 10^{-5}$ \citep[e.g. ZIMPOL,][]{zimpol}, but this is accomplished by sacrificing either spatial resolution or temporal cadence (or both). With the advent of the National Solar Observatory's Daniel K. Inouye Solar Telescope (DKIST), and other upcoming observatories, this boundary will be pushed, but we will still hover around $\sigma_{\rm n}=10^{-4}$ for moderate to good spatial resolutions\footnote{See the requirements for the polarimetric accuracy of DKIST data at http://dkist.nso.edu/science/intro}.
Assessing the validity of a diagnostic tool such as WFA has little value if noise is not accounted for. Therefore, we added random noise to all of the synthetic Stokes profiles, and re-evaluated the accuracy of the WFA under more realistic conditions. For each set of Stokes profiles, we added 200 different realizations of the noise at the $\sigma_{\rm n}=10^{-3}$ and the $\sigma_{\rm n}=10^{-4}$ levels (relative to the continuum intensity), and calculated the $3\sigma$ uncertainty in the inferred magnetic field for each of the two signal-to-noise (S/N) scenarios.

\begin{figure}
\includegraphics[angle=0,scale=.49]{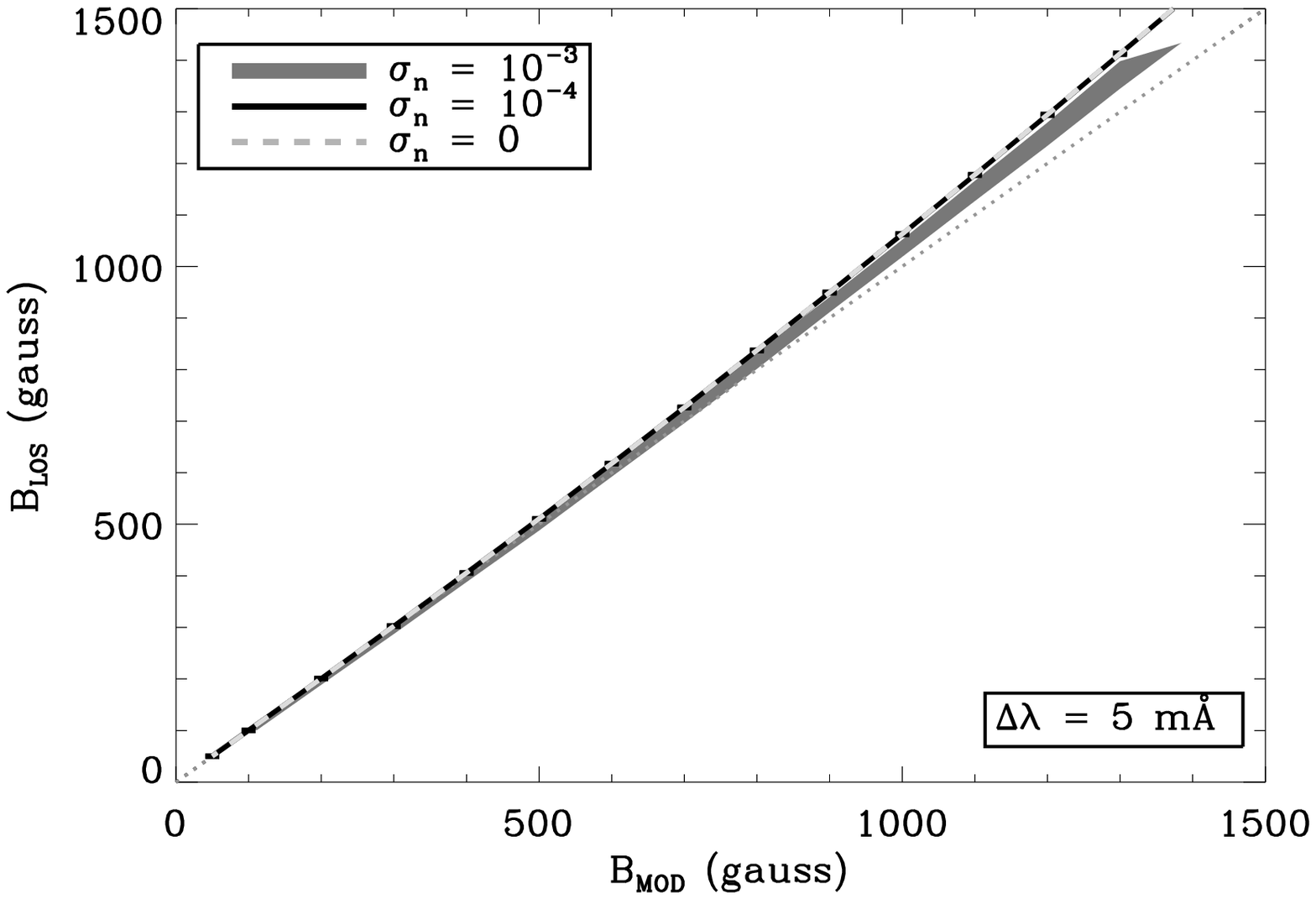}
\includegraphics[angle=0,scale=.49]{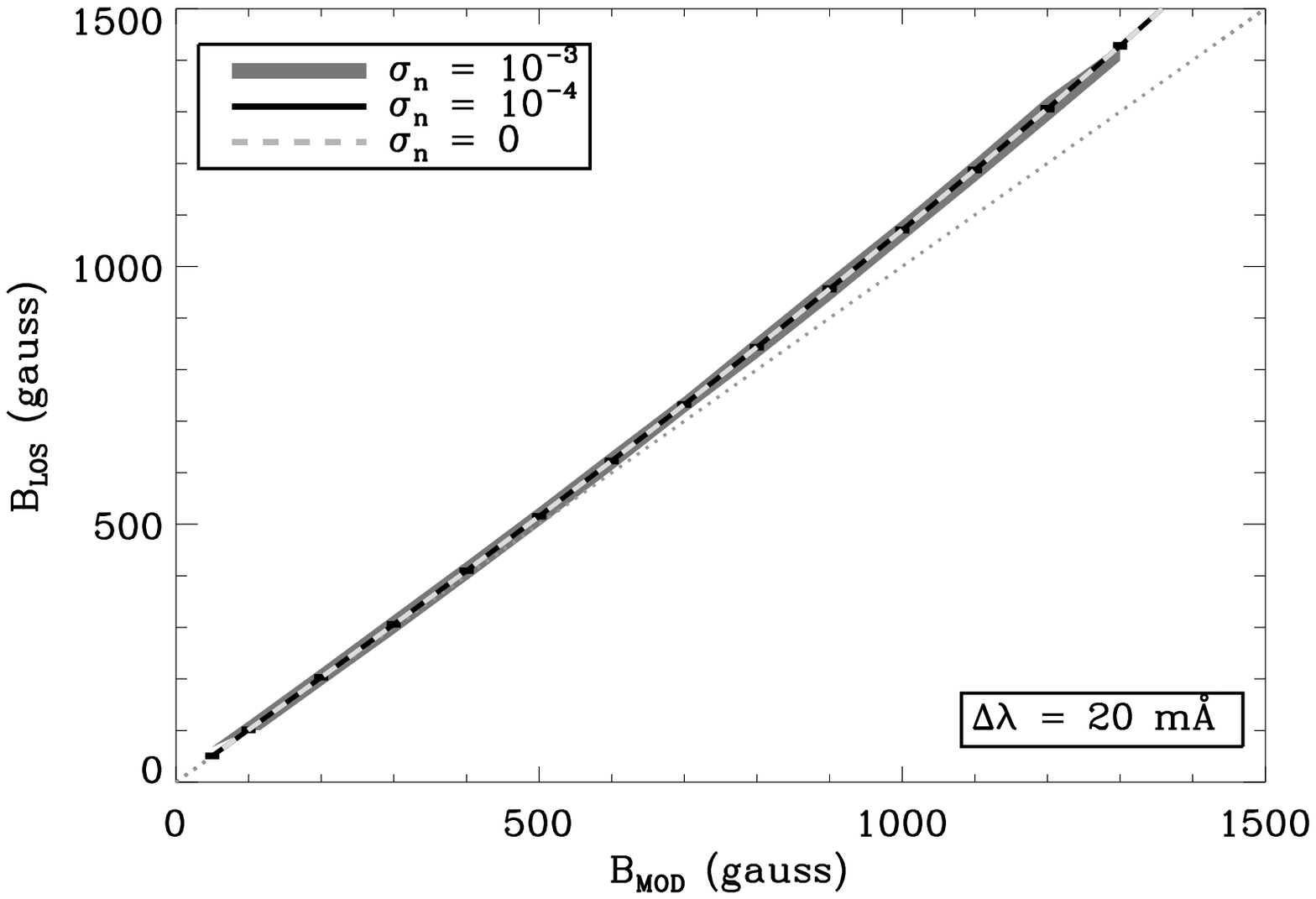}\\
\includegraphics[angle=0,scale=.49]{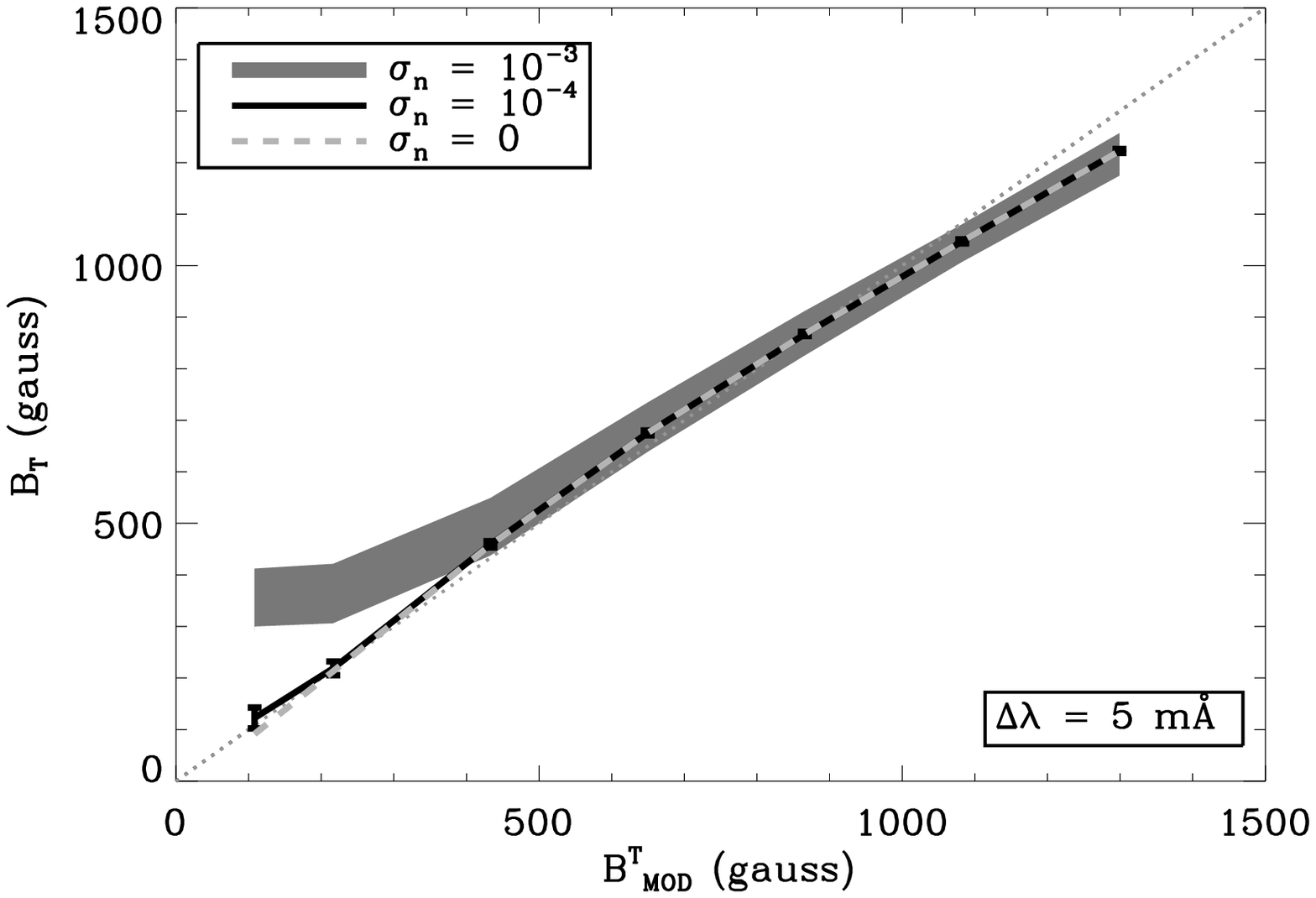}
\includegraphics[angle=0,scale=.49]{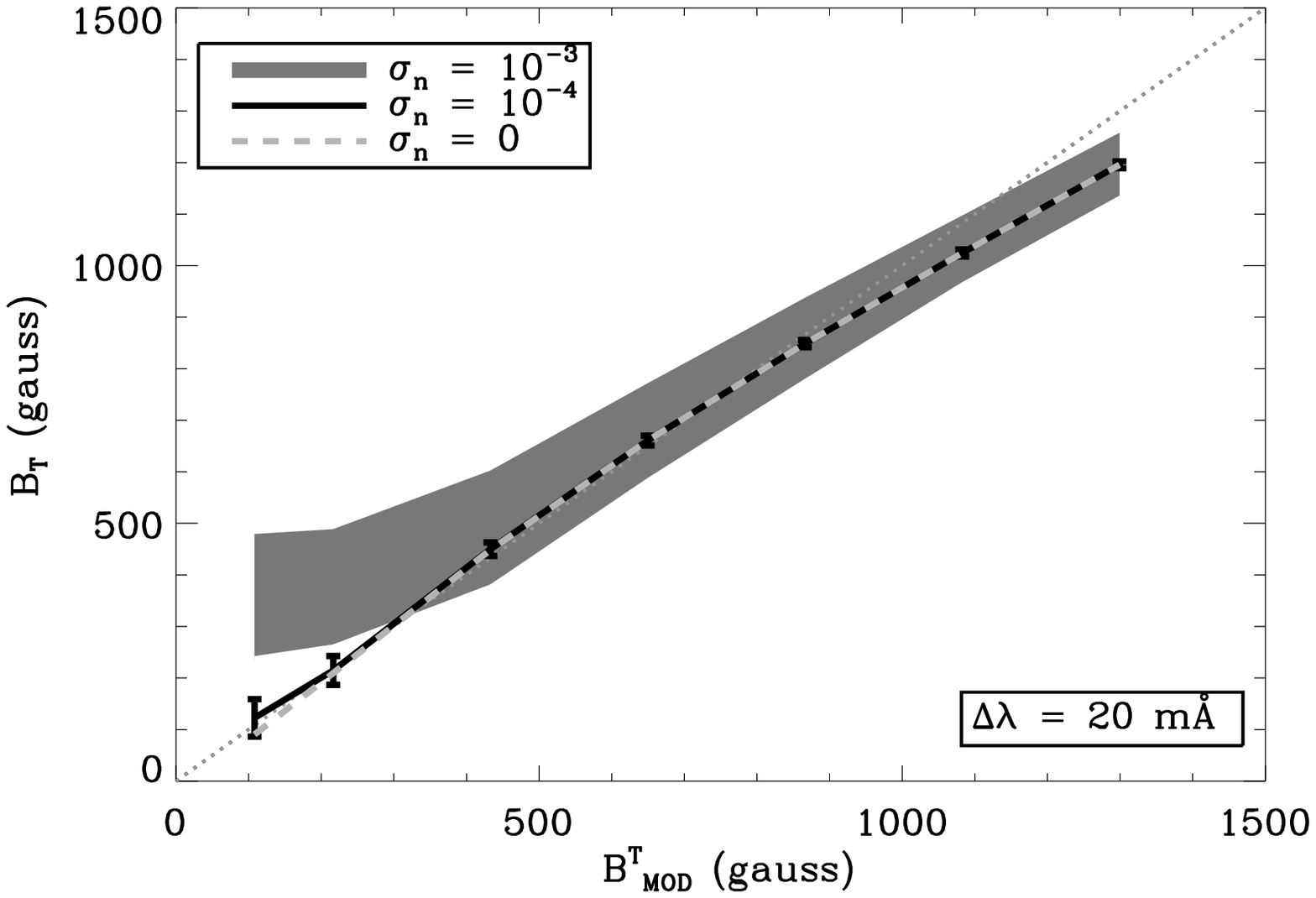}\\
\includegraphics[angle=0,scale=.49]{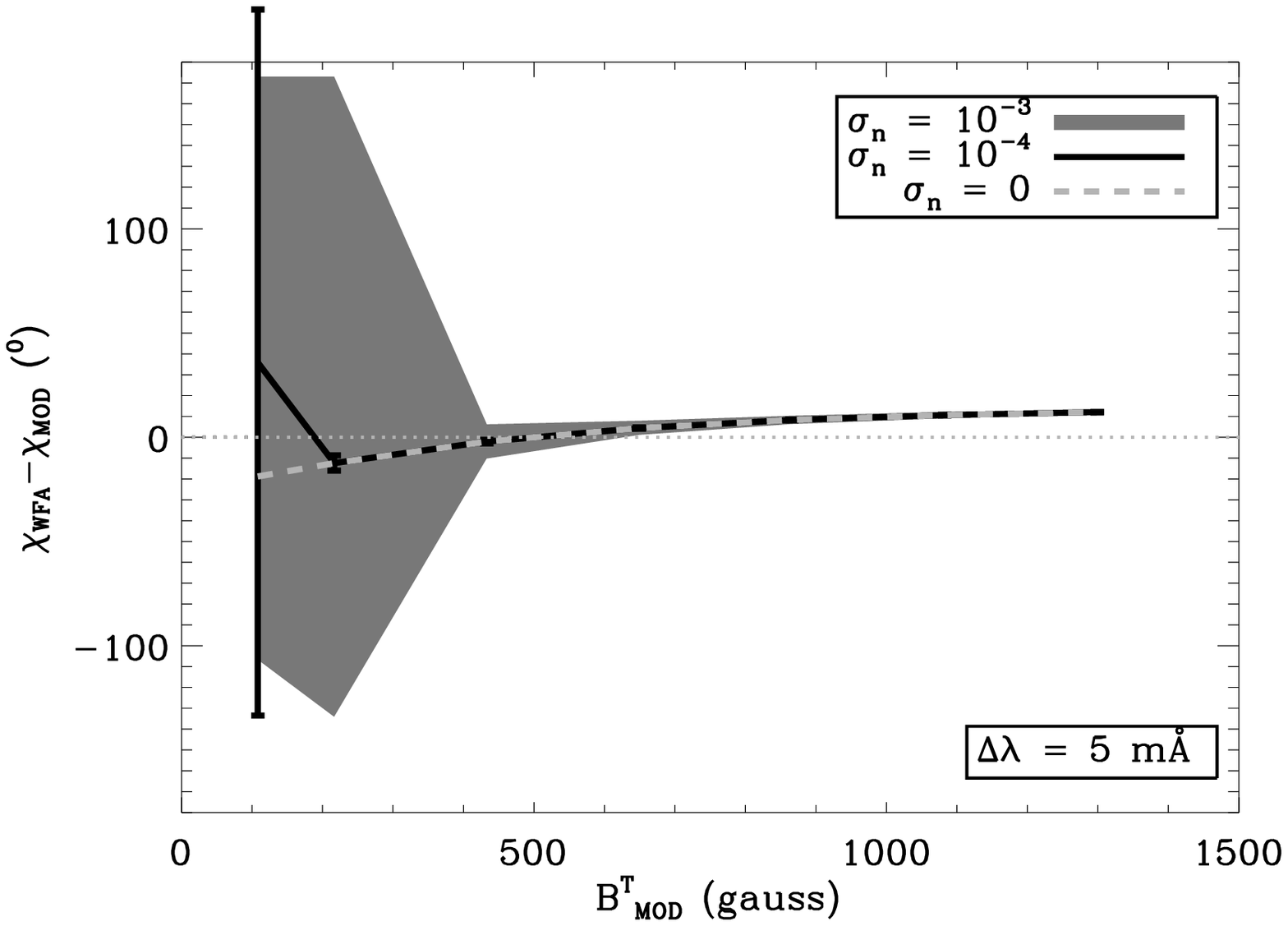}
\includegraphics[angle=0,scale=.49]{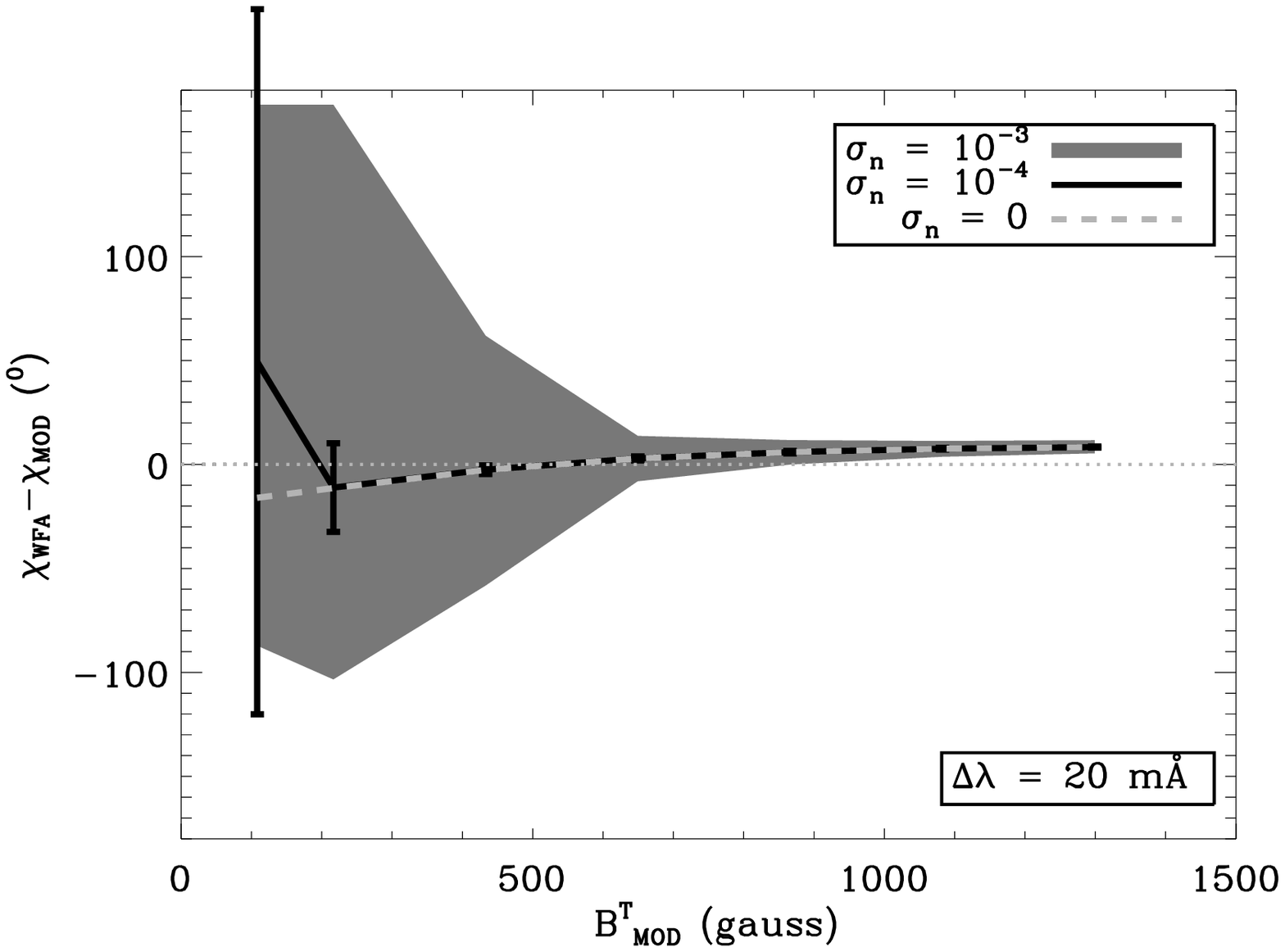}
\caption{Effects of noise and spectral sampling on the WFA inferences. The shaded gray areas represent the 3$\sigma$ uncertainty in the inferred components of the magnetic field for the case of a noise level of $\sigma_{\rm n}=10^{-3}$ (relative to the continuum intensity), whilst the solid black lines with error bars correspond to the case of $\sigma_{\rm n}=10^{-4}$.  The noiseless reference is depicted by the dashed grey line in all plots. The results on the left side were obtained from spectra computed on a 5 m\AA\ grid, whilst the panels on the right resulted from spectra with a 20 m\AA\ sampling.
The top row shows the errors in the retrieval of $B_{\rm LOS}$ from the core of the spectral line ($\lambda_0\pm 0.25$\AA), in the case of a constant vertical magnetic field. The middle row presents the uncertainties in the retrieval of $B_{\rm T}$ (estimated from the wing-core boundary) for the case of a constant, yet inclined magnetic field ($\theta=60^{\circ}$). The bottom row shows the $3\sigma$ uncertainties in the retrieval of the magnetic field azimuth when noise is added to the observations.\label{fig:noise}}
\end{figure}

The top left panel of Fig. \ref{fig:noise} shows the line-of-sight magnetic field inferred from synthetic profiles with different S/N levels for the case of a constant vertical magnetic field. The dashed gray line represents the noiseless reference case.
For a noise level of $\sigma_{\rm n}=10^{-4}$ (solid black line), the $3\sigma$ error bars are negligible in comparison to the systematics, and the results follow the noiseless inference. 
The shaded grey area shows the $3\sigma$ spread in the inferred field strengths for a noise level of $\sigma_{\rm n}=10^{-3}$. The uncertainty due to the noise is of similar magnitude to the systematic errors of the WFA method but the inferred values present a systematic under-estimation with respect to the noiseless scenario (dashed grey line). This departure from the noiseless case is a consequence of the formulation in Eq. \ref{eq:wfablosfit}, where the quadratic term in the denominator results in an additive noise effect and a systematic under-estimate of the line-of-sight field strength. Albeit closer to the ideal result in this particular instance, this effect typically biases the inferred field strengths to lower values and worsens with increasing noise and increasing number of wavelength samples. 
By contrast, the top right panel shows the WFA inferences for the same noise levels and same spectra, but sampled every 20 m\AA\ instead of every 5 m\AA. Because we only use a quarter of the wavelength samples, the additive effect of the noise in the denominator of Eq. \ref{eq:wfablosfit} is less severe, and the average inferences for the $\sigma_{\rm n}=10^{-3}$ S/N case fall on top of the noiseless results. From a practical perspective, this shows a case for binning observations along the spectral direction in order to reduce the noise before applying equation \ref{eq:wfablosfit}.

\noindent The middle left panel of Fig. \ref{fig:noise} shows the inferred transverse component of the magnetic field for the two S/N scenarios, as well as the noiseless case for the sake of comparison (dashed grey line). As described in section \ref{sec:horizontal}, Eq. \ref{eq:btransfit} was applied in the $\lambda-\lambda_0 = [-0.4,-0.1]$~\AA\ wavelength range. At low field strengths, the noise dominates over the polarization signals and the transverse magnetic field is over-estimated. This effect is much more significant in the case of $\sigma_{\rm n}=10^{-3}$ (shaded grey area) than for $\sigma_{\rm n}=10^{-4}$ (solid black line). Inferring $B_{\rm T}$ from Eq. \ref{eq:btransfit} requires computing the absolute value of the wavelength derivative of Stokes $I$, as well as the total linear polarization, which is a sum of the squared linear polarization profiles. Both operations amplify rather than cancel the noise, rendering it the dominant component in the calculation for transverse fields below $\sim 450$~G. Above this threshold, the error introduced by the noise still results in overestimating the field by a few percent, for the range of model field strengths tested in this work. For a noise level of $\sigma_{\rm n}=10^{-4}$, the effect is only noticeable below $B_{\rm T} \sim 200$G, with the systematic biases of WFA becoming the dominant source of error above that value. 

\noindent If we reduce the number of wavelength samples by a factor of four, keeping the same noise levels, the $3\sigma$ spread in the inferred values becomes larger, particularly for the $\sigma_{\rm n}=10^{-3}$ case, in the middle right panel of the figure.

The inference of the azimuth of the magnetic field in the POS is also affected by the presence of noise in the observations. The bottom left panel of Fig. \ref{fig:noise} shows the difference between the inferred azimuth and the model value as a function of the transverse field strength. While the dashed line is the reference case (no noise), the shaded grey area shows the uncertainty resulting from $\sigma_{\rm n}=10^{-3}$ and the black solid line represents the case of $\sigma_{\rm n}=10^{-4}$. The noise disproportionally affects the $\chi_{\rm WFA}$ inferences for weak transverse fields. Below $B_{\rm T}\sim 400$~G, the azimuth retrieved from observations with a $\sigma_{\rm n}=10^{-3}$ is completely unreliable. For a $\sigma_{\rm n}=10^{-4}$, this threshold is lowered to $\sim 200$~G, below which, the inference departs from the noiseless case. 
In contrast to the case of the inference of $B_{\rm LOS}$, the reduction in the number of wavelength samples renders the azimuth inference less reliable in the presence of noise. For a noise level of $\sigma_{\rm n}=10^{-3}$, a 20 m\AA\ sampling results in very large uncertainties of the WFA inferences for transverse fields with strengths below $\sim 650$~G. The $\sigma_{\rm n}=10^{-4}$ case, on the other hand, is a lot more forgiving, setting the lower threshold around $\sim 300$G. This is due to the formulation of Eq. \ref{eq:azimuthfit}, where noise cancels out in the summatory terms of the numerator and the denominator, whereby a larger number of wavelength samples results in better noise cancellation.

\section{Discussion and Conclusions}\label{sec:discussion}
 
The computational overhead associated to spectral line inversions of chromospheric lines is such that these methods are currently not viable tools for near-real-time data processing or for the study of very large datasets. The weak field approximation is an easy and fast way of extracting magnetic field information from spectropolarimetric observations, and is often used in the context of quick-look data products. In this work, we assess the range of validity of the WFA for the analysis of the chromospheric Ca {\sc ii} 8542 \AA\ line.

By using the WFA on synthetic Stokes profiles generated with the spectral line synthesis code NICOLE, we were able to test the limits of its applicability. One interesting diagnostic capability of the Ca {\sc ii} 8542 \AA\ line comes from the fact that its core forms in the chromosphere while its wings are purely photospheric in origin \citep{quinteronoda}. This can be exploited to extract the depth dependence of the magnetic properties in the Sun's atmosphere.
When evaluated in the core of the spectral line ($\lambda_0 \pm 0.25$\AA), the WFA for the line of sight magnetic field delivers results that are accurate within 10\% of the solution, for field strengths up to 1200~G. This spectral range probes chromospheric layers at an average optical depth of $\log \tau \sim -5.3$. The wing of the line ($\lambda-\lambda_0 = [-2,-1]$~\AA), on the other hand, can be used to probe the line-of-sight photospheric magnetic field around optical depth $\log \tau \sim -1.4$, and the accuracy is remarkable for the entire range of magnetic field strengths, gradients and values of the spectral smearing tested in Section \ref{sec:gradientb}.
The presence of a moderate velocity gradient, albeit having visible imprints on the spectral profiles, has very little impact on the WFA inferences. However, the presence of a velocity discontinuity (of 3 kms$^{-1}$ in magnitude) around the height of the formation of the core of the line, hinders the accurate determination of the line-of-sight magnetic field.

The inference of the chromospheric transverse field is done using Eq. \ref{eq:btranswing}, evaluated in the core-wing boundary ($\lambda - \lambda_0 = [-0.4, -0.1]$). This is a compromise between omitting the bulk of the scattering polarization signatures near the line core while still avoiding the outer photospheric wings.
The inferred transverse field strengths are accurate typically within 10\% of the solution, however they get worse for purely transverse fields ($\theta = 90^{\circ}$). Also, the larger the spectral smearing, the less accurate the solution, probably owing to the limitations of Eq. \ref{eq:btransfit} when incorporating information from the inner core of the line.

\noindent The selected wavelength range for the application of Eq. \ref{eq:btranswing} will not completely avoid scattering polarization signatures.
Although its signals become the dominant source of linear polarization close to the limb, they are still relatively weak across the majority of the solar disk. If the Zeeman signals are large enough ($B_{\rm T}>200$~G), the WFA inference still yields results within $20\%$ of the model value for heliocentric angles within $30^{\circ}$ of the limb. A detailed assessment of the impact of scattering polarization signals on spectral line inversions and WFA inferences from Ca {\sc ii} 8542 \AA\ will be the topic of future work.

The azimuth of the magnetic field in the plane of the sky is extracted from Eq. \ref{eq:azimuthfit}. The inference of the azimuth, $\chi_{\rm WFA}$, from the wing of the line reflects the orientation of the field at photospheric layers. Albeit less accurate (with errors of up to $\sim 10^{\circ}$ depending on the field strength and its inclination angle),  the wing-core boundary should be better suited to study the magnetic field azimuth in the chromosphere. The reason for the discrepancies between the models and the WFA inferences in the latter case is the presence of magneto-optical effects, which substantially affects the linear polarization profiles, and exerts a much bigger impact on the core than the wings of the spectral line.

 There are many ways of computing linear least-squares regressions in order to find the most probable slope to fit observational data. Different methods, however, yield mathematically different results \citep{isobe, feigelson}. The approaches used to derive equations \ref{eq:wfablosfit}, \ref{eq:btransfit} and \ref{eq:azimuthfit} have been chosen, not because they provide optimal results, but because they are some of the most common approaches used in the literature. It is beyond the scope of this work to evaluate all the available regression methods.

Noise is an unavoidable fact of life, and in particular of spectropolarimetric observations. By adding noise to the synthetic Stokes profiles at the $\sigma_{\rm n}= 10^{-3}$ and $\sigma_{\rm n}= 10^{-4}$ levels (relative to the continuum intensity), we were able to evaluate the effect this had on the WFA inferences.

\noindent We conclude that a noise level of $\sigma_{\rm n}= 10^{-3}$, while acceptable for the extraction of $B_{\rm LOS}$, is still too large for inferences of the horizontal component of the field strength and its azimuth in the plane of the sky using the WFA: $B_{\rm T}$ is severely over-estimated and the uncertainty in the azimuth is extremely large for field strengths below $\sim450$~G. 
There is a big qualitative leap from the $\sigma_{\rm n}= 10^{-3}$ to the $\sigma_{\rm n}= 10^{-4}$ case, the latter resulting in more accurate estimates of $B_{\rm T}$ and $\chi_{\rm WFA}$ above $\sim 200$~G. 

\noindent Because of the formulation of Eqs. \ref{eq:wfablosfit} and \ref{eq:btransfit}, the presence of noise in the observations leads to a systematic bias that increases with the number of wavelength samples. This results in an under-estimation of $B_{\rm LOS}$ and an over-estimation of $B_{\rm T}$. The inference of the azimuth, on the other hand, benefits from the larger number of wavelength samples as noise tends to cancel out in Eq. \ref{eq:azimuthfit}.

 Noise in the observations does not affect exclusively the WFA, since \cite{jaime_model} report similar results for spectral line inversions of the Ca {\sc ii} 8542 \AA\ line emerging from a magneto-hydrodynamic simulation. Although the Monte-Carlo-type procedure used on synthetic spectra in Section \ref{sec:noise} can be applied to observations, Bayesian methods are also powerful tools to extract valuable information of the limitations derived from noise in the data \citep{asensio}.

A word of caution is in order. The details of the spectral profiles of the Ca {\sc ii} 8542 \AA\ line depend on how the forward calculation is done. Both the choice of the semi-empirical model atmosphere in which the line is synthesized, as well as the spectral line synthesis code used for the radiative transfer calculations, will affect the shape of the spectral profiles. Parallel tests carried out with Han Uitenbroek's RH\footnote{http://www4.nso.edu/staff/uitenbr/rh.html} code indicate that, while the exact quantitative results depend on the code used, the general behavior is consistent among the two.

 All in all, the weak field approximation remains a valid method to extract chromospheric magnetic field information from the Ca {\sc ii} 8542 \AA\ line in certain field strength regimes. But when applying it, one has to be weary of its range of validity, not only in regards to the {\em weak field} regime per se, but also by assessing the presence of magneto-optical effects, scattering polarization signatures, and the biases introduced by photon noise.

The author would like to thank Tanaus\'u del Pino Alem\'an and Roberto Casini for valuable discussions on the effects of scattering polarization and for their radiative transfer code used for the calculations in Section \ref{sec:scatt_pol}. The author would also like to thank the anonymous referee of this paper for valuable remarks that resulted in some additional sections to this manuscript. The National Center for Atmospheric Research is sponsored by the National Science Foundation.




\end{document}